\documentclass{tlp} 

\usepackage{epsf}
\usepackage{aopmath} 
\usepackage{amssymb}
\usepackage{proof}
\usepackage{euscript}
\usepackage{amsfonts}   
\usepackage{latexsym}

\newcount\PLv\newcount\PLw\newcount\PLx\newcount\PLy\newdimen\PLyy\newdimen\PLX
\newbox\PLdot \setbox\PLdot\hbox{\tiny.} \def\scl{.08} 
\def\PLot#1{\PLx`#1\advance\PLx-42\PLy\PLx\PLv\PLx\divide\PLy9\PLw\PLy\multiply
\PLw9\advance\PLx-\PLw\advance\PLx-4\PLy-\PLy\advance\PLy4\PLX=\the\PLx pt
\advance\PLyy\the\PLy pt\wd\PLdot=\scl\PLX\raise\scl\PLyy\copy\PLdot}
\def\draw#1{\ifx#1\end\let\next=\relax\else\PLot#1\let\next=\draw\fi\next}

\def\invamp{\hbox{\PLyy=70pt\draw :::;DMV_gqppyyyyyooooxxxnnwvlutkjaWNE=5-./9%
9:::CCCC:::99/..--544=EENWWaajjjkktttttttNNNVVVVVVVV\end \hskip4pt}}
\newbox\iabox\setbox\iabox\invamp \def\Invamp{\copy\iabox}

\newcommand{\with}{\,\&\,}
\newcommand{\lolli}{\mathbin{-\hspace{-0.70mm}\circ}}
\newcommand{\lollo}{\mathbin{\circ\hspace{-0.70mm}-}}

\newcommand{\para}{\mathrel{\Invamp}}

\newcommand{\all}{\top}
\newcommand{\one}{ {\bf 1} }

\newcommand{\arrowupdown}[4]{\setbox0=\hbox{$\ {}^{#2}\ $}
  \setbox1=\hbox{$\longrightarrow$}
  \ifdim\wd0<\wd1\setbox0=\box1\else\relax\fi
  {#1}\,\mathop{\hbox to \wd0{\rightarrowfill}}\limits^{#2}_{#3}\,{#4}
}

\newcommand{\arrowdown}[3]{\setbox0=\hbox{$\ {}_{#2}\ $}
  \setbox1=\hbox{$\longrightarrow$}
  \ifdim\wd0<\wd1\setbox0=\box1\else\relax\fi
  {#1}\,\mathop{\hbox to \wd0{\rightarrowfill}}\limits_{#2}\,{#3}
}

\newcommand\Arrow[3]{\setbox0=\hbox{$\ {}^{#2}\ $}
  \setbox1=\hbox{$\Rightarrow$}
  \loop\setbox1=\hbox{=\kern-0.3em\unhbox1}\ifdim\wd1<\wd0\repeat
  \hbox{${#1}\,\mathop{\box1}\limits^{#2\,}{#3}$}
}

\newcommand{\rhssep}{\mid\!\mid}

\newtheorem {es}{Example}

\newenvironment{example}{\begin{es}}{\hspace*{\fill}$\Box$\end{es}}

\newcommand {\calA} {\mbox{${\cal A}$}} 
\newcommand {\calB} {\mbox{${\cal B}$}}
\newcommand {\calC} {\mbox{${\cal C}$}}
\newcommand {\calD} {\mbox{${\cal D}$}}

\newcommand {\calI} {\mbox{${\cal I}$}}

\newcommand {\calK} {\mbox{${\cal K}$}} 

\newcommand {\calM} {\mbox{${\cal M}$}}

\newcommand {\calP} {\mbox{${\cal P}$}}

\newcommand{\Nat}{\mbox{\cal N}}

\newcommand{\smallunion}{\cup}
\newcommand{\myunion}{\bigcup}

\newcommand{\compos}{\circ}

\newcommand{\dedLO} [2] {{#1}\Rightarrow{#2}}
\newcommand{\dedLOo} [2] {{#1}\Rightarrow_\one{#2}}

\renewcommand {\seq} [6] {{#1};{#2}[{#3}]\rightarrow_{#4}{#5\,\rhssep\,#6}}

\newcommand {\cand}{\wedge}
\newcommand {\clor}{\vee}

\newcommand {\eset} {\emptyset}

\newcommand {\tpo} {T_P^\one}
\newcommand {\fpo} {F_\one(P)}

\newcommand {\itp} [1] {T_P\!\!\uparrow_{#1}}
\newcommand {\itpo} [1] {T_P^\one\!\!\uparrow_{#1}}
\newcommand {\isp} [1] {S_P\!\!\uparrow_{#1}}

\newcommand {\itt} [1] {T\!\!\uparrow_{#1}}
\newcommand {\Un} [2] {\myunion_{{#1}=1}^{#2}}

\renewcommand {\eps} {\epsilon}

\newcommand {\opo} {O_\one(P)}

\newcommand {\deno} [1] {[\![{#1}]\!]_\one}

\newcommand {\preq} {\preccurlyeq}

\newcommand {\STPo} {S_P^\one}

\newcommand {\asato} {\asat_\one}

\newcommand {\msc}[1] {\widehat{#1}}

\newcommand {\LOone} {LO${}_\one$\,}
\newcommand {\LOonet} {LO{\Large ${}_\one$}\,}
\newcommand {\LOonews} {LO${}_\one$}

\newtheorem{definition}{Definition}[section]

\newcommand{\ms}[1]{\widehat{#1}}

\newcommand{\formula}[1]{{\bf #1}}

\renewcommand{\vec}[1]{ {\mathbf{#1}} }

\newcommand{\sat}{\models}
\newcommand{\osat}{\sat}

\newcommand {\val} [3] {#1 \sat #2 [#3]}
\newcommand {\valo} [3] {#1 \osat #2 [#3]}
\newcommand {\symbvalo} [3] {#1 \asat #2 [#3]}

\newcommand{\den}[1]{[\![{#1}]\!]}
\newcommand{\tuple}[1]{\langle{#1}\rangle}

\newcommand{\leftlolli}{\lollo}
\newcommand{\asat}{\Vdash}
\newcommand{\STP}{S_P}
\newcommand{\mlub}[2]{#1\bullet #2}
\newcommand{\Add}{\alpha}
\newcommand{\Rem}{\rho}

\newcommand{\linc} {LC_\Sigma}

\newcommand{\tpa} {T_P^{\#}}
\newcommand{\TPA} {\tpa}
\newcommand{\tpao} [1] {T_P^{\#}\!\!\uparrow_{#1}}

\newcommand{\trad} [1] {\lceil {#1} \rceil}

\newcommand{\cimp} {\leftarrow}

\newcommand{\true} {\mbox {\tt tt}}

\newcommand{\dtp} {T^{dlp}_P}
\newcommand{\lotp} {T^{lo}_P}
\newcommand{\tpatr} {T^{\#}_{\lceil P\rceil}}
\newcommand{\lotpo} [1] {\lotp\!\!\uparrow_{#1}}
\newcommand{\lotptr} [1] {T^{lo}_{\lceil P\rceil}}

\newcommand{\dhbp} {DHB_P}

\newcommand{\dosp} {O^{dlp}_P}
\newcommand{\dfs} {F^{dlp}_P}
\newcommand{\lofs} {F^{lo}_P}

\newcommand{\lfp} {\mathit {lfp}}
\newcommand{\gnd} {\mathit {gnd}}

\title[Fixpoint Semantics for  Linear Logic Programs]
  {An Effective Fixpoint Semantics for  Linear Logic Programs}
\author[M. Bozzano, G. Delzanno and M. Martelli]
{MARCO BOZZANO, GIORGIO DELZANNO and MAURIZIO MARTELLI\\
Dipartimento di Informatica e Scienze dell'Informazione\\
Universit\`a di Genova\\
Via Dodecaneso 35, 16146 Genova - Italy \\
\email{\{bozzano,giorgio,martelli\}@disi.unige.it}
}
\pagerange{\pageref{firstpage}--\pageref{lastpage}}
\volume{}
\jdate{}
\pubyear{}

\begin{document}
\maketitle
\label{firstpage}
\begin{abstract}
In this paper we investigate the theoretical foundation of a new 
bottom-up semantics for linear logic programs, and more precisely 
for the fragment of LinLog \cite{And92} that
consists of the language LO \cite{AP91a}
enriched with the constant $\one$.
We use {\em constraints} to symbolically and finitely 
represent possibly infinite collections  of provable goals.
We define a fixpoint semantics based on a new operator in the style
of $T_P$ working over constraints.
An application of the fixpoint operator can be computed algorithmically.
As sufficient conditions for termination, we show that
the fixpoint computation is guaranteed to converge for propositional LO.
To our knowledge, this is the first attempt to define an effective
fixpoint semantics for linear logic programs. 
As an application of our framework, we also present a formal investigation
of the relations between LO and Disjunctive Logic Programming \cite{MRL91}. 
Using an approach based on abstract interpretation, we show that
DLP fixpoint semantics can be viewed as an abstraction of our semantics for LO.
We prove that the resulting abstraction is {\em correct} and {\em complete}
\cite{CC77,GR97b} for an interesting class of LO programs encoding Petri Nets.
\end{abstract}
\section{Introduction}
In recent years a number of fragments of linear logic \cite{Gir87} have been 
proposed as a logical foundation for extensions of
logic programming\cite{Mil95}.
Several new programming languages like {LO} \cite{AP91a}, LinLog \cite{And92},
ACL \cite{KY95}, Lolli \cite{HM94}, and Lygon \cite{HP94} have been proposed 
with the aim of enriching traditional logic programming languages like Prolog 
with a well-founded notion of state and with aspects of concurrency.
The operational semantics of this class of languages is given via 
a sequent-calculi presentation of the corresponding fragment of linear logic.
Special classes of proofs like the {\em focusing} proofs of \cite{And92} 
and the {\em uniform} proofs of \cite{Mil96} allow us to restrict 
our attention to {\em cut-free}, {\em goal-driven} proof systems that are 
complete with respect to provability in linear logic.
These presentations of linear logic are the natural counterpart of the traditional 
{\em top-down} operational semantics of logic programs.

In this paper we investigate an alternative operational semantics 
for the fragment of linear logic underlying the language LO \cite{AP91a}, and
its proper extension with the constant $\one$. Both languages can be seen as
fragments of LinLog \cite{And92}, which is a presentation of full linear
logic. Throughout the paper, we will simply refer to these two fragments as
LO and \LOonews.
The reason we selected these fragments is that 
we were looking for a relatively simple linear logic language 
with a uniform-proof presentation, state-based computations and 
aspects of concurrency.  
Considering both LO and its extension with the constant $\one$ will help us 
to formally classify the different expressive power of linear logic connectives
like $\para$, $\with$, $\top$, and $\one$ when incorporated into a logic programming setting.
In practice, {LO} has been successfully applied to model 
concurrent object-oriented languages \cite{AP91a},
and multi-agent coordination languages based on the Linda model \cite{And96}.

The operational semantics we propose consists of a {\em goal-independent}
{\em bottom-up} evaluation of programs.
Specifically, given an {LO} program $P$ our aim is to compute a finite 
representation of the set of goals that are provable from $P$.
There are several reasons to look at this problem.
First of all, as discussed in \cite{HW98}, 
the bottom-up evaluation of programs is the key ingredient
for all applications where it is difficult or impossible to specify a given 
goal in advance. Examples are active (constraint) databases, agent-based 
systems and genetic algorithms. 
Recent results connecting verification techniques and 
semantics of logic programs \cite{DP99} show that bottom-up evaluation can be used 
to automatically check  {\em properties} (specified in temporal logic like CTL) 
of the original program. 
In this paper will go further showing that the {\em provability} relation in 
logic programming languages like LO can be used to naturally 
express verification problems for Petri Nets-like models of concurrent systems.
Finally,  a formal definition of the bottom-up semantics 
can be useful for studying equivalence, compositionality and 
abstract interpretation, as for traditional logic programs 
\cite{BGLM94,GDL95}.

Technically, our contributions are as follows.
We first consider a formulation of {LO} with $\para$, $\lolli$, $\with$ and $\all$.
Following the semantic framework of (constraint) logic programming 
\cite{GDL95,JM94}, we formulate the bottom-up evaluation procedure in two
steps. 
We first define what one could call a {\em ground} semantics 
via a {\em fixpoint} operator $T_P$ defined over an extended notion 
of Herbrand interpretation consisting of {\em multisets} of atomic formulas. 
This way, we capture the {\em uniformity} of {LO}-provability, according to which 
compound goals must be completely decomposed into atomic goals before 
program clauses can be applied. 
Due to the structure of the LO proof system, 
already in the propositional case there 
are infinitely many provable  multisets of atomic formulas. 
In fact, {LO}-provability enjoys the following property.
If a multiset of goals $\Delta$ is provable in $P$, then 
any $\Delta'$ such that $\Delta$ is a sub-multiset of
$\Delta'$ is provable in $P$.   
To circumvent this problem, we order the interpretations according 
to the multiset inclusion relation of their elements and we define a
new operator $\STP$ that computes only the {\em minimal} (w.r.t. 
multiset inclusion) provable multisets. 
Dickson's Lemma \cite{Dic13} ensures the termination of the 
fixpoint computation based on $\STP$ for propositional {LO} programs.
Interestingly, this result is an instance of the general decidability 
results for model checking of infinite-state systems given in \cite{ACJT96,FS01}.

The decidability of propositional  provability shows that LO
is not as interesting as one could expect from a state-oriented extension 
of the logic programming paradigm.
Specifically, LO does not provide a natural way to {\em count} resources.
This feature can be introduced by 
a slight extension of LO in which we add unit clauses defined
via the constant $\one$.
The resulting language, namely \LOone, can be viewed as a first step
towards more complex languages based on linear logic like LinLog
\cite{And92}.
As we show in this paper, \LOone allows to model more sophisticated models of 
concurrent systems than LO, e.g., in \LOone it is possible to model Petri Nets with 
{\em transfer arcs}.   
Adding the constant $\one$  breaks down the decidability of provability in propositional {LO}.
Despite this negative result, it is still possible to define 
an effective $\STPo$ operator for \LOone.
For this purpose, 
as {\em symbolic} representation of potentially infinite sets of contexts, 
we choose a special class of {\em linear constraints} defined over 
variables that {\em count} resources.
This abstract domain generalizes the domain used for LO:
the latter can be represented as the subclass of constraints with 
no equalities.
Though for the new operator we cannot guarantee that the fixpoint can be
reached after finitely many steps, this connection allows us to apply
techniques developed in model checking for infinite-state systems 
(see e.g. \cite{BGP97,DP99,HHW97})
and abstract interpretation \cite{CH78} 
to compute approximations of the fixpoint of $\STPo$.

In this paper we limit ourselves to the study of the propositional
case that, as shown in \cite{APC97}, can be viewed as the target of 
a possible abstract interpretation of a first-order program.
To our knowledge, this is the first attempt of defining an effective
fixpoint semantics for linear logic programs.

Our semantic framework can also be used as a tool to compare the relative
strength of different logic programming extensions. 
As an application, we shall present a detailed comparison between LO and Disjunctive Logic
Programming (DLP).
Though DLP has been introduced in order to represent `uncertain' beliefs,
a closer look at its formal definition reveals very interesting
connections with the paradigm of linear logic programming:
both DLP and LO programs extend Horn programs allowing clauses with
{\em multiple} heads. 
In fact, in DLP we find  clauses of the form 
$$
p(X)\vee q(X) \leftarrow r(X)\wedge t(X),
$$
whereas in LO we find clauses of the form 
$$
p(X)\para q(X) \leftlolli r(X)\with t(X).
$$
To understand the differences, we must look at the operational 
semantics of DLP programs.
In DLP, a resolution step is extended so as to work over positive 
clauses ({\em sets}/{\em disjunctions} of facts). 
Implicit {\em contraction} steps are applied  over the selected clause.
On the contrary,
being in a sub-structural logic in which contraction is forbidden,
we know that LO resolution behaves as {\em multiset} rewriting.
Following the bottom-up approach that we pursue in this paper, we will exploit
the classical framework of abstract interpretation to formally compare the
two languages. Technically, we first specialize our fixpoint semantics to a
{\em flat} fragment of propositional
LO (i.e., arbitrary nesting of connectives in goals is forbidden) which
directly corresponds to DLP as defined in \cite{MRL91}. Then,
by using an abstract-interpretation-based approach,
we exhibit  a Galois connection between the semantic domains of DLP and LO,
and we show that the semantics of DLP programs can be described
as an abstraction of the semantics of LO programs.
Using the theory of abstract interpretation and 
the concept of {\em complete abstraction} \cite{CC77,GR97b} we
discuss the quality of the resulting abstraction.
This view of DLP as an abstraction of LO is appealing for several reasons.
First of all, it opens the possibility of using techniques developed for DLP 
for the analysis of LO programs.  
Furthermore, it shows that the paradigm of DLP could have unexpected
applications
as a framework  to reason about properties of Petri Nets, a well-know
formalism for concurrent computations \cite{KM69}.
In fact, as we will prove formally in the paper, DLP represents a
{\em complete} abstract domain for LO programs that encode Petri Nets. 
\paragraph*{Plan of the paper.}
After introducing some notations in Section \ref{prelim},
in Section \ref{lo} we recall the main features of {LO} \cite{AP91a}.
In Section \ref{ground} we introduce the so-called {\em ground} semantics,
via the $T_P$ operator,
and prove that the least fixpoint of $T_P$ characterizes the operational 
semantics of an LO program.
In Section \ref{nonground} we reformulate LO semantics by means of
the {\em symbolic} $\STP$ operator, and we relate it to $T_P$.
In Section \ref{groundo} we consider an extended fragment of LO
with the constant $\one$,
extending the notion of satisfiability given in Section \ref{ground} and
introducing an operator $\tpo$.
In Section \ref{nongroundo} we introduce a symbolic operator $\STPo$
for the extended fragment, and we discuss its algorithmic implementation
in Section \ref{nongroundo_computing}. As an application of our framework, in
Section \ref{dlplo} and Section \ref{comparison} we investigate the relations 
between LO and DLP, and in Section \ref{petri} we investigate the relations with Petri Nets.
Finally, in Section \ref{related} and Section \ref{conclusions} we discuss
related works and conclusions.

This paper is an extended version of the papers \cite{BDM00a,BDM00b}.
\section{Preliminaries}
\label{prelim}
In this paper we will extensively use operations on multisets.
We will consider a fixed signature, i.e a finite set of propositional
symbols, $\Sigma=\{a_1,\ldots,a_n\}$. Multisets over $\Sigma$
will be hereafter called {\em facts}, and symbolically
noted as $\calA,\calB,\calC,\ldots$.
A multiset with (possibly duplicated) elements $b_1,\ldots,b_m\in\Sigma$
will be simply indicated as $\{b_1,\ldots,b_m\}$, overloading the usual
notation for sets.

A fact $\calA$ is uniquely determined by a finite map 
$Occ:\Sigma\rightarrow\Nat$ such that $Occ_{\cal A}(a_i)$ is the number of occurrences of 
$a_i$ in $\calA$.
Facts are ordered according to the {\em multiset inclusion} 
relation $\preccurlyeq$ defined as follows: $\calA\preccurlyeq\calB$ 
if and only if $Occ_{\cal A}(a_i)\leq Occ_{\cal B}(a_i)$ for $i:1,\ldots,n$.
The {\em empty} multiset is denoted $\eps$ and is such that 
$Occ_\eps(a_i)=0$ for $i:1,\ldots,n$, and $\eps\preccurlyeq\calA$ 
for any $\calA$. 
The {\em multiset union} $\calA,\calB$ (alternatively $\calA+\calB$ when `,' is ambiguous) 
of two facts $\calA$ and $\calB$ is such that 
$Occ_{{\cal A},{\cal B}}(a_i)=Occ_{\cal A}(a_i)+Occ_{\cal B}(a_i)$ for $i:1,\ldots,n$.
The {\em multiset difference}  $\calA\setminus\calB$ is such that
$Occ_{{\cal A}\setminus{\cal B}}(a_i)=max(0,Occ_{\cal A}(a_i)-Occ_{\cal B}(a_i))$ 
for $i:1,\ldots,n$.
We define a special operation $\mlub{}{}$ to compute the {\em least upper 
bound} of two facts with respect to $\preccurlyeq$. Namely,
$\mlub{\calA}{\calB}$ is such that
$Occ_{\mlub{\cal A}{\cal B}}(a_i)=max(Occ_{\cal A}(a_i),Occ_{\cal B}(a_i))$ for 
$i:1,\ldots,n$.
Finally, we will use the notation $\calA^n$, where $n$ is a natural number, to
indicate $\calA+\ldots+\calA$ ($n$ times).

In the rest of the paper we will use $\Delta,\Theta,\ldots$ to denote
multisets of possibly compound formulas. Given two multisets 
$\Delta$ and $\Theta$, $\Delta\preccurlyeq\Theta$ indicates multiset inclusion
and 
$\Delta,\Theta$ multiset union,
as before, and $\Delta,\{G\}$ is written simply $\Delta,G$.
In the following, a {\em context} will denote a multiset of goal-formulas
(a {\em fact} is a context in which every formula is atomic).
Given a linear disjunction of atomic formulas $H=a_1\para\ldots\para a_n$,  
we introduce the notation $\ms{H}$ to denote the multiset $a_1,\ldots,a_n$.

Finally, let $T:\calI\rightarrow\calI$ be an operator defined over a complete lattice 
 $\tuple{\calI,\sqsubseteq}$.  
We define $\itt{0}=\eset$, where $\eset$ is the
bottom element, $\itt{k+1}=T(\itt{k})$ for all $k\geq 0$, and
$\itt{\omega}=\bigsqcup_{k=0}^{\infty}\itt{k}$, 
where $\bigsqcup$ is the least upper bound w.r.t. $\sqsubseteq$.
Furthermore, we use $\lfp(T)$ to denote the {\em least fixpoint} of $T$. 
\section{The Programming Language {LO}}
\label{lo}
{LO} \cite{AP91a} is a logic programming language based on linear logic. 
Its mathematical foundations lie on a proof-theoretical presentation of 
a fragment of linear logic defined over the linear connectives 
$\leftlolli$ ({\em linear implication}), $\with$ ({\em additive conjunction}),
$\para$ ({\em multiplicative disjunction}), and the constant $\all$
({\em additive identity}). 
In the propositional case {LO} consists of the following class of formulas:

$$
\begin{array}{l}
\formula{D}\ ::=\ \formula{A}_1\para\ \ldots\para\ \formula{A}_n\ \leftlolli\ \formula{G}\ \ |\ \ \formula{D}\ \with\ 
\formula{D}\\
[\medskipamount]
\formula{G}\ ::=\ \formula{G}\ \para\ \formula{G}\ \ |\ \formula{G}\ \with\ \formula{G}\ \ |\ \ 
\formula{A}\ \ |\ \ \all
\end{array}
$$
Here $\formula{A}_1,\ldots,\formula{A}_n$ and $\formula{A}$ 
range over propositional symbols from a fixed signature $\Sigma$. 
$\formula{G}$-formulas correspond to {\em goals} to be evaluated in a given program. 
$\formula{D}$-formulas correspond to multiple-headed {\em program clauses}. 
An {LO} program is a $\formula{D}$-formula.  
Let $P$ be the program $C_1\with\ldots\with C_n$.
The execution of  a multiset of $\formula{G}$-formulas $G_1,\ldots,G_k$ 
in $P$ corresponds to a goal-driven proof for the two-sided {LO}-sequent 
$$P\Rightarrow G_1,\ldots,G_k.$$
The {LO}-sequent $P\Rightarrow G_1,\ldots,G_k$ is an abbreviation for 
the following two-sided linear logic sequent: 
$$!C_1,\ldots,!C_n\ \rightarrow\   G_1,\ldots,G_k.$$
The formula $!F$ on the left-hand side of a sequent indicates that $F$ can 
be used in a proof an arbitrary number of times.
This implies that an {LO}-Program can be viewed also as a 
{\em set of reusable clauses}.
According to this view,  the operational semantics of {LO} is given via 
the {\em uniform} (goal-driven)  proof system defined in Figure \ref{system_for_LO}. 
In Figure \ref{system_for_LO}, $P$ is a set of implicational clauses,
${\cal A}$ denotes a multiset of atomic formulas,
whereas $\Delta$ denotes a multiset of $\formula{G}$-formulas.
A sequent is provable if all branches of its proof tree
terminate with instances of the $\all_r$ axiom.
The proof system of Figure \ref{system_for_LO} is a specialization of more general 
uniform proof systems for linear logic like Andreoli's focusing proofs \cite{And92}, 
and Forum \cite{Mil96}. 
\begin{figure*}
$$
\begin{array}{c}
\infer[\all_r]
{P\Rightarrow \all,\Delta}
{}
\ \ \ \ 
\infer[\para_r]
{P\Rightarrow G_1\para G_2,\Delta}
{P\Rightarrow G_1,G_2,\Delta}
\ \ \ \ 
\infer[\with_r]
{P\Rightarrow G_1\with G_2,\Delta}
{P\Rightarrow G_1,\Delta &
 P\Rightarrow G_2,\Delta}
\\
\\
\infer[bc\ \ (H\leftlolli G\ \in\ P)]
{P\Rightarrow \ms{H}+{\cal A}}
{P\Rightarrow G,{\cal A} & }
\end{array}
$$
\caption{A proof system for {LO}}
\label{system_for_LO}
\end{figure*}
The rule {\em bc} denotes a backchaining (resolution) step 
($\ms{H}$ is the multiset consisting of the literals in 
the disjunction $H$, see Section \ref{prelim}).
Note that {\em bc} can be executed only if the right-hand side of 
the current {LO} sequent consists of atomic formulas.
Thus, {LO} clauses behave like {\em multiset} rewriting rules. 
{LO} clauses having the following form
$$a_1\para\ \ldots\para\  a_n\leftlolli \all$$
play the same role as the unit clauses of Horn programs.
In fact, a backchaining step over such a clause 
leads to {\em success} independently of the current context $\calA$,
as shown in the following scheme:
$$ 
\begin{array}{c}
\infer[bc]
{P\Rightarrow a_1,\ldots,a_n,\calA}
{\infer[\all_r]
  {P\Rightarrow \all,\calA}{} 
} 
\medskip\\
provided\ \ \ a_1\para\ \ldots\para \ a_n\leftlolli\all\in P
\end{array}
$$
This observation leads us to the following property (we recall that 
$\preccurlyeq$ is the sub-multiset relation).
\begin{proposition}
\label{monotonicity}
Given an {LO} program $P$ and two contexts $\Delta,\Delta'$ 
such that $\Delta\preccurlyeq\Delta'$, if 
$P\Rightarrow \Delta$ then $P\Rightarrow \Delta'$.
\end{proposition}
\begin{proof}
By simple induction on the structure of LO proofs.
\end{proof}
This property is the key point in our analysis of the operational 
behavior of {LO}. It states that the $weakening$ rule is $admissible$ in LO.
Thus, LO can be viewed as an {\em affine} fragment of linear logic.
Note that weakening and contraction are both admissible on the left hand side 
(i.e. on the program part) of LO sequents.
\begin{example}
\label{LOex}
Let $P$ be the {LO} program consisting of the clauses
$$\begin{array}{l}
  1.\ \ a\lollo b\para c\\
  2.\ \ b\lollo(d\para e)\with f\\
  3.\ \ c\para d\lollo\all\\
  4.\ \ e\para e\lollo b\para c\\
  5.\ \ c\para f\lollo\all\\
\end{array}$$
and consider an initial goal $e,e$. 
A proof for this goal is shown in Figure \ref{LOproof},
where we have denoted by $bc^{(i)}$ the application of the backchaining rule
over clause number $i$ of $P$. The proof proceeds as follows.
Using clause 4., to prove $e,e$ we have to prove
$b\para c$, which, by {LO} $\para_r$ rule, reduces to prove $b,c$.
At this point we can backchain over clause 2., and we get the new goal
$(d\para e)\with f,c$. By applying $\with_r$ rule, we get
two separate goals $d\para e,c$ and $f,c$. The first, after a reduction
via $\para_r$ rule, is provable by means of clause (axiom) 3., while the latter
is provable directly by clause (axiom) 5.
Note that $\all$ succeeds in a non-empty context (i.e. containing $e$) in
the left branch.
A similar proof shows that the goal $a$ is also provable from $P$.
\begin{figure*}
\centering
$$
   \infer[bc^{(4)}]{P\Rightarrow e,e}
  {\infer[\para_r]{P\Rightarrow b\para c}
  {\infer[bc^{(2)}]{P\Rightarrow b,c}
  {\infer[\with_r]{P\Rightarrow (d\para e)\with f,c}
  {\infer[\para_r]{P\Rightarrow d\para e,c}
  {\infer[bc^{(3)}]{P\Rightarrow d,e,c}
  {\infer[\all_r]{P\Rightarrow e,\all}{}}}
    &
   \infer[bc^{(5)}]{P\Rightarrow f,c}
  {\infer[\all_r]{P\Rightarrow \all}{}}}
  }}}
$$
\caption{An LO proof for the goal $e,e$ in the program of Example \ref{LOex}}
\label{LOproof}
\end{figure*}
By Proposition \ref{monotonicity}, provability of $e,e$ and $a$
implies provability of any multiset of goals $e,e,\Delta$ and $a,\Delta$, for
every context $\Delta$.
\end{example}

We conclude this Section with the definition
of the following induction measure on LO goals, which we will later
need in proofs.
\begin{definition}
\label{inddef}
Given a goal $G$, the induction measure $m(G)$ is defined according to the
following rules: $m(A) = 0$ for every atomic formula $A$;
$m(\all) = 0$; $m(G_1\with G_2) = m(G_1\para G_2) = m(G_1) + m(G_2)$ + 1.
The induction measure extends to contexts by defining
$m(G_1,\ldots,G_n)=m(G_1)+\ldots+m(G_n)$.
\end{definition}
\section{A Bottom-up Semantics for {LO}}
\label{ground}
The proof-theoretical semantics of {LO} corresponds to the {\em top-down} 
operational semantics based on resolution for traditional logic 
programming languages like Prolog. 
Formally, we define the operational {\em top-down} semantics of an {LO}
program $P$ as follows:
$$ 
O(P)=\{\calA\ |\ \calA\ is\ a\ fact\ and\ \dedLO{P}{\calA}\ is\ provable\}
$$
Note that the information on provable {\em facts} from a given program $P$
is all we need to decide
whether a general goal (with possible nesting of connectives) is provable from
$P$ or not. This is a consequence
of the {\em focusing} property \cite{And92} of LO
provability, which ensures that provability of a compound goal can always be
reduced to provability of a finite set of atomic multisets.
In a similar way, in Prolog 
the standard bottom-up semantics is defined as a set of {\em atoms}, while
in general {\em conjunctions} of atoms are allowed in clause bodies.

In this paper we are interested in finding a suitable definition of
{\em bottom-up} 
semantics that can be used as an alternative operational semantics for {LO}.
More precisely, given an {LO} program $P$ we would like to define a procedure
to compute
all goal formulas $G$ such that $G$ is provable from $P$.
This procedure should enjoy the usual properties of classical bottom-up
semantics, in particular its definition should be based on an {\em effective}
fixpoint operator (i.e. at least every single step must
be finitely computable),
and it should be {\em goal-independent}. As usual, goal independence is
achieved by searching for proofs starting from the axioms (the unit clauses of
Section \ref{lo}) and accumulating goals which can be proved by applying
program clauses to the current interpretation.
As for the operational semantics, we can limit ourselves to goal formulas 
consisting of multisets of atomic formulas, without any loss of generality.  
In the rest of the paper we will always consider propositional {LO} programs
defined over a {\em finite} set of propositional symbols $\Sigma$.
We give the following definitions.
\begin{definition}[Herbrand base $B_P$]
Given a pro\-positional {LO} program $P$ defined over $\Sigma$,
the Herbrand base of $P$, denoted $B_P$, is given by
  $$ B_P = \{\calA \ |\  \calA\ \hbox{is a multiset (fact) over $\Sigma$}\}.$$
\end{definition}
\begin{definition}[Herbrand interpretation]
We say that $I\subseteq B_P$ is a Herbrand interpretation.
Herbrand interpretations form a complete
lattice $\tuple{\calD,\subseteq}$ with respect to set inclusion,
where $\calD=\calP(B_P)$.
\end{definition}
Before introducing the formal definition of the {\em ground} bot\-tom-up semantics, 
we need to define a notion of satisfiability of a context  $\Delta$ in a 
given interpretation $I$.
For this purpose, we introduce the judgment $\val{I}{\Delta}{\calA}$.
The need for this judgment, with respect to the familiar logic programming setting
\cite{GDL95}, is motivated by the arbitrary nesting of connectives in LO
clause bodies,
which is not allowed in traditional presentations of (constraint)
logic programs.
In $\val{I}{\Delta}{\calA}$, $\calA$ should be read as an {\em output} fact
such that 
$\calA+\Delta$ is valid in $I$.
This notion of {\em satisfiability} is modeled according to the right-introduction 
rules of the connectives. The notion of output fact $\calA$ will simplify the 
presentation of the algorithmic version of the judgment which we will present in Section \ref{nonground}.
\begin{definition}[Satisfiability]
Let $I$ be a Herbrand interpretation, then $\sat$ is defined as follows:
$$\begin{array}{l} 
  \val{I}{\all,\Delta}{\calA'}\ \mbox{for\ any\ fact}\ \calA';\\
  [\medskipamount]
  \val{I}\calA{\calA'}\ \mbox{if}\ \calA+\calA' \in I;\\
  [\medskipamount]
   \val{I}{G_1\para G_2,\Delta}{\calA}\ \mbox{if}\ \val{I}{G_1,G_2,\Delta}{\calA};\\
  [\medskipamount]
  \val{I}{G_1\with G_2,\Delta}{\calA}\ \mbox{if}\ \val{I}{G_1,\Delta}{\calA}\ 
  \mbox{and}
    \ \val{I}{G_2,\Delta}{\calA}.
\end{array}$$
\end{definition}
The relation $\models$ satisfies the following properties.
\begin{lemma}
\label{movlemma}\label{satlemma}\label{finitary_lemma}
For any interpretations $I,J$,
context $\Delta$, and fact $\calA$,
\begin{enumerate}
\item[i)] $\val{I}{\Delta}{\calA}\ if\ and\ only\ if\ \val{I}{\Delta,\calA}{\eps}$;
\item[ii)] if $I\subseteq J$ and $\val{I}{\Delta}{\calA}$ then $\val{J}{\Delta}{\calA}$;
\item[iii)]
given a chain of interpretations $I_1\subseteq I_2\subseteq\ldots$, 
if  $\val{\Un{i}{\infty}I_i}{\Delta}{\calA}$ 
then there exists $k$ s.t. $\val{I_k}{\Delta}{\calA}$.
\end{enumerate}
\end{lemma}
\begin{proof}
The proof of $i)$ and $ii)$ is by simple induction.
The proof of $iii)$ is by (complete) induction on $m(\Delta)$ (see Definition
\ref{inddef}).
\begin{itemize}
  \item[-] If $\Delta=\all,\Delta'$, then, no matter which $k$ you choose,
    $\val{I_k}{\all,\Delta'}{\calA}$;
  \item[-] if $\Delta$ is a fact, then 
    $\val{\Un{i}{\infty}I_i}{\Delta}{\calA}$ means
    $\Delta,\calA\in\Un{i}{\infty}I_i$, which in turn implies that there
    exists $k$ such that
    $\Delta,\calA\in I_k$, therefore $\val{I_k}{\Delta}{\calA}$;
  \item[-] if $\Delta=G_1\with G_2,\Delta'$, then by inductive hypothesis
    there exist $k_1$ and $k_2$ s.t. $\val{I_{k_1}}{G_1,\Delta'}{\calA}$ and
    $\val{I_{k_2}}{G_2,\Delta'}{\calA}$. Therefore, if $k=max\{k_1,k_2\}$, by
    $ii)$ we
    have that $\val{I_k}{G_1,\Delta'}{\calA}$
    and $\val{I_k}{G_2,\Delta'}{\calA}$, therefore
    $\val{I_k}{G_1\with G_2,\Delta'}{\calA}$, i.e.
    $\val{I_k}{\Delta}{\calA}$ as required;
  \item[-] the $\para$-case follows by a straightforward
    application of the inductive hypothesis.
\end{itemize}
\end{proof}
We now come to the definition of the fixpoint operator $T_P$.
\begin{definition}[Fixpoint operator $T_P$]
Given a program $P$ and an interpretation $I$,
the operator $T_P$ is defined as follows:
  $$ T_P(I)=\{\ms{H}+\calA\ |\ H\lollo G\in P,\ \val{I}{G}{\calA}\}. $$
\end{definition}
The following property holds.
\begin{proposition}\label{montheor}
For every program $P$, $T_p$ is monotonic and continuous over the lattice
$\tuple{\calD,\subseteq}$.
\end{proposition}
\begin{proof}
{\em Monotonicity}. Immediate from $T_P$ definition and 
Lemma \ref{satlemma} $ii)$.\\
{\em Continuity}. We prove that $T_P$ is finitary. 
Namely, given an increasing chain of interpretations $I_1\subseteq I_2\subseteq\ldots$,
$T_P$ is finitary if 
$T_P(\Un{i}{\infty}I_i)\subseteq \Un{i}{\infty}{T_P(I_i)}$.
We simply need to show that if
$\val{T_P(\Un{i}{\infty}I_i)}{\Delta}{\eps}$ then\ there\ exists\ $k$ \ such\ that\ 
         $\val{T_P(I_k)}{\Delta}{\eps})$. 
	 The proof is by induction on $m(\Delta)$.
    \begin{itemize}
      \item[-] If $\Delta=\all,\Delta'$, then, no matter which $k$ you choose,
        $\val{T_P(I_k)}{\all,\Delta'}{\eps}$;
      \item[-] if $\Delta$ is a fact and $\Delta\in
        T_P(\Un{i}{\infty}I_i)$,
        then, by definition of $T_P$, there exist a fact $\calA$
        and a clause $A_1\para\ldots\para A_n\lollo G\in P$, such that
        $\val{\Un{i}{\infty}I_i}{G}{\calA}$
        and $\Delta=A_1,\ldots,A_n,\calA$.
        Lemma \ref{finitary_lemma} $iii)$ implies that $\exists k. \val{I_k}{G}{\calA}$,
        therefore, again by definition of $T_P$,
        $\val{T_P(I_k)}{A_1,\ldots,A_n,\calA}{\eps}$, i.e.
        $\val{T_P(I_k)}{\Delta}{\eps}$ as required;
      \item[-] if $\Delta=G_1\with G_2,\Delta'$, then by inductive hypothesis,
        there exist $k_1$ and $k_2$ s.t.
        $\val{T_P(I_{k_1})}{G_1,\Delta'}{\eps}$
        and $\val{T_P(I_{k_2})}{G_2,\Delta'}{\eps}$.
        Then, if $k=max\{k_1,k_2\}$,
        by Lemma \ref{satlemma} $ii)$  we have that
        $\val{T_P(I_k)}{G_1,\Delta'}{\eps}$ and
        $\val{T_P(I_k)}{G_2,\Delta'}{\eps}$. This implies
        $\val{T_P(I_k)}{G_1\with G_2,\Delta'}{\eps}$, i.e.
        $\val{T_P(I_k)}{\Delta}{\eps}$ as required;
      \item[-] the $\para$-case follows by a straightforward application of the
        inductive hypothesis.
    \end{itemize}
\end{proof}
Monotonicity and continuity of the $T_P$ operator imply, by Tarski's Theorem,
that $\lfp(T_P)=\itp{\omega}$.

Following \cite{Llo87}, we define the {\em fixpoint semantics} 
$F(P)$ of an {LO} program $P$  as the least fixpoint of $T_P$, 
namely $F(P)=\lfp(T_P)$. 
Intuitively, $T_P(I)$ is the set of {\em immediate logical consequences} 
of the program $P$ and of the facts in $I$. 
In fact, if we define $P_I$ as the program $\{A\leftlolli\all\ |\ \ms{A}\in I\}$,
the definition of $T_P$ can be viewed as the following instance 
of the {\em cut} rule of linear logic:
$$
\infer[cut]
{!P,!P_I \rightarrow H,\calA}
{
!P,G\rightarrow H & 
!P_I\rightarrow G,\calA
}
$$
Using the notation used for {LO}-sequents we obtain the following rule:
$$
\infer[cut]
{P \cup P_I \Rightarrow H,\calA}
{
P\Rightarrow H\leftlolli G & 
P_I\Rightarrow G,\calA
}
$$
Note that, since  $H\leftlolli G\in P$, the sequent 
$P\Rightarrow H\leftlolli G$ is always provable in linear logic.
According to this view, $F(P)$ characterizes the set of 
{\em logical consequences} of a program $P$.

The fixpoint semantics is sound and complete with respect
to the operational semantics as stated in the following theorem.
\begin{theorem}[Soundness and Completeness]
\label{fixpoint_operational}
For every {LO} program $P$, $F(P)=O(P)$.
\end{theorem}
\begin{proof}
\begin{description}
  \item[$i)\ F(P)\subseteq O(P)$.] We prove that for every $k$ and
    context $\Delta$,
    if $\val{\itp{k}}{\Delta}{\eps}$ then $\dedLO{P}{\Delta}$.
    The proof is by (complete) induction on $\bar{m}(\itp{k},\Delta)$, where
    $\bar{m}$ is an induction measure defined by
    $\bar{m}(\itp{k},\Delta)=\langle k,m(\Delta)\rangle$, and
    $\langle k,m\rangle < \langle k',m'\rangle$ if and only if
    ($k<k'$) or ($k=k'$ and $m<m'$) (lexicographic ordering).
    \begin{itemize}
      \item[-] If $\Delta=\all,\Delta'$, the conclusion is immediate;
      \item[-] if $\Delta$ is a fact, then $\Delta\in \itp{k}$, so
        that $\itp{k}\neq\eset$ and $k>0$. By definition of $T_P$ we have
        that there exist a fact $\calA$ and a clause
        $A_1\para\ldots\para A_n\lollo G\in P$, such that
        $\val{\itp{k-1}}{G}{\calA}$ and $\Delta=A_1,\ldots,A_n,\calA$.
        By Lemma \ref{movlemma} $i)$ we have that
        $\val{\itp{k-1}}{G,\calA}{\eps}$, and then, by inductive
        hypothesis, $\dedLO{P}{G,\calA}$, therefore by LO $bc$ rule,
        $\dedLO{P}{A_1,\ldots,A_n,\calA}$, i.e., $\dedLO{P}{\Delta}$;
      \item[-] if $\Delta=G_1\with G_2,\Delta'$ then by inductive hypothesis
        $\dedLO{P}{G_1,\Delta'}$ and $\dedLO{P}{G_2,\Delta'}$, therefore
        $\dedLO{P}{G_1\with G_2,\Delta'}$ by LO $\with_r$ rule;
      \item[-] the $\para$-case follows by a straightforward
        application of the inductive hypothesis.
    \end{itemize}
  \item[$ii)\ O(P)\subseteq F(P)$.] We prove that for every context $\Delta$
    if $\dedLO{P}{\Delta}$ then there exists $k$ such that
    $\val{\itp{k}}{\Delta}{\eps}$
    by induction on the structure of the LO proof.
    \begin{itemize}
      \item[-] If the proof ends with an application of $\all_r$, then the
        conclusion is immediate;
      \item[-] if the proof ends with an application of the $bc$ rule, then
        $\Delta=A_1,\ldots,A_n,\calA$, where $A_1,\ldots,A_n$ are atomic
        formulas, and there exists a clause
        $A_1\para\ldots\para A_n\lollo G\in P$.
        For the uniformity of LO proofs, we can
        suppose $\calA$ to be a fact. By inductive hypothesis,
        we have that there exists $k$ such that
        $\val{\itp{k}}{G,\calA}{\eps}$, then, by Lemma \ref{movlemma} $i)$,
        $\val{\itp{k}}{G}{\calA}$, which, by definition of $T_P$, in turn
        implies that $A_1,\ldots,A_n,\calA\in T_P({\itp{k}})=\itp{k+1}$,
        therefore $\val{\itp{k+1}}{A_1,\ldots,A_n,\calA}{\eps}$,
        i.e., $\val{\itp{k+1}}{\Delta}{\eps}$;
      \item[-] if the proof ends with an application of the $\with_r$ rule,
        then $\Delta=G_1\with G_2,\Delta'$ and, by inductive hypothesis,
        there exist $k_1$ and $k_2$ such that
        $\val{\itp{k_1}}{G_1,\Delta'}{\eps}$ and
        $\val{\itp{k_2}}{G_2,\Delta'}{\eps}$. Then, if $k=max\{k_1,k_2\}$
        we have, by Lemma \ref{satlemma} $ii)$, that
        $\val{\itp{k}}{G_1,\Delta'}{\eps}$ and
        $\val{\itp{k}}{G_2,\Delta'}{\eps}$, therefore
        $\val{\itp{k}}{G_1\with G_2,\Delta'}{\eps}$, i.e.
        $\val{\itp{k}}{\Delta}{\eps}$;
      \item[-] the $\para$-case follows by a straightforward
        application of the inductive hypothesis.
    \end{itemize}
\end{description}
\end{proof}
We note that it is also possible to define a {\em model-theoretic} semantics
(as for classical logic programming \cite{GDL95})
based on the notion of {\em least model} with respect to a given
class of models and partial order relation. In our setting, the partial order
relation is simply set inclusion, while models are exactly
Herbrand interpretations which satisfy program clauses, i.e., $I$ is a model
of $P$ if and only if for every clause $H\lollo G\in P$ and for every fact
$\calA$, $$ \val{I}{G}{\calA} \ \hbox{implies}\  \val{I}{H}{\calA}. $$
It turns out that the operational, fixpoint and model-theor\-etic semantics are
all equivalent. We omit details.
Finally, we also note that these semantics can be proved equivalent to the
{\em phase semantics} for LO given in \cite{AP91a}.

\section{An Effective Semantics for LO}
\label{nonground}
The operator $T_P$ defined in the previous section
does not enjoy one of the crucial properties we required for our
bottom-up semantics, namely its definition is {\em not} effective.
As an example, take the program $P$ consisting of the clause $a\leftlolli\all$.
Then, $T_P(\emptyset)$ is the set of all multisets with {\em at least}
one occurrence of $a$, which is an {\em infinite} set. 
In other words, $T_P(\emptyset)=\{\calB\ |\  a\preccurlyeq \calB\ \}$,
where $\preccurlyeq$ is the multiset inclusion relation of Section
\ref{prelim}.
In order to compute {\em effectively} one step of $T_P$, we have to find
a {\em finite} representation of potentially infinite sets of facts 
(in the terminology of \cite{ACJT96}, a {\em constraint system}). 
The previous example suggests us that a provable fact {\em \calA} may be used to 
{\em implicitly} represent the ideal generated by $\calA$, i.e., 
the subset of $B_P$ defined as follows:
$$
\den{\calA}=\{\calB\ |\ \calA\preccurlyeq\calB\}.
$$
We extend the definition of $\den{\cdot}$ to sets of facts as follows: 
$\den{I}=\bigcup_{{\cal A}\in I} \den{\calA}$. 
Based on this idea, we define an {\em abstract} Herbrand base 
where  we handle every single fact $\calA$ as a representative element
for $\den{\calA}$ (note that in the semantics of Section \ref{ground} 
the denotation of a fact $\calA$ is $\calA$ itself!).
\begin{definition}[Abstract Herbrand Inter\-pret\-ation]
\label{interpretation_lattice}
The lattice $\tuple{\calI,\sqsubseteq}$ of abstract Herbrand interpretations is
defined as follows:
\begin{itemize} 
\item[-] $\calI=\calP(B_P)/\simeq$ where $I\simeq J$ if and only 
if $\den{I}=\den{J}$;
\item[-] $[I]_\simeq\sqsubseteq [J]_\simeq$ if and only if for all $\calB\in I$ there exists $\calA\in J$ 
such that $\calA\preccurlyeq\calB$;
\item[-] the bottom element is the empty set $\emptyset$,
      the top element is the $\simeq$-equivalence class of the singleton 
 $\{\eps\}$ ($\eps$=empty multiset,
 $\eps\preccurlyeq \calA$ for any $\calA\in B_P$);
\item[-] the least upper bound $I\sqcup J$ is the $\simeq$-equivalence class of $I\cup J$.
\end{itemize} 
\end{definition}
The equivalence $\simeq$ allows us to reason modulo {\em redundancies}. 
For instance, any $\calA$ is redundant in $\{\eps,\calA\}$, which, in fact, is
equivalent to $\{\eps\}$. 
It is important to note that to compare two ideals we 
simply need to compare their generators w.r.t. the multiset inclusion relation
$\preccurlyeq$. 
Thus, given a {\em finite} set of facts we can always remove all redundancies
using a polynomial number of comparisons.
\paragraph*{Notation.}
For the sake of simplicity, in the rest of the paper we will identify an 
interpretation $I$ with its class $[I]_\simeq$.
Furthermore, note that if
$\calA\preccurlyeq\calB$, then $\den{\calB}\subseteq\den{\calA}$. 
In contrast,  if $I$ and $J$ are two interpretations and $I\sqsubseteq J$ 
then $\den{I}\subseteq\den{J}$.
\\
\mbox{}

The two relations $\preccurlyeq$ and $\sqsubseteq$ are {\em well-quasi orderings}
\cite{ACJT96,FS01}, as stated in Proposition \ref{dickson} and 
Corollary \ref{termination_lemma} below.
This property is the key point of our idea. 
In fact, it will allow us to prove that the computation of  
the least fixpoint of the {\em symbolic} formulation of the
operator $T_P$ (working on abstract Herbrand interpretations)
is guaranteed to terminate on every input {LO} program. 
\begin{proposition}[Dickson's Lemma \cite{Dic13}]\label{dickson}
Let $A_1 A_2 \ldots$ be an infinite sequence of 
multisets over the finite alphabet $\Sigma$. 
Then there exist two indices $i$ and $j$ such that $i<j$ and 
$A_i\preccurlyeq A_j$.
\end{proposition}
Following \cite{ACJT96}, by definition of $\sqsubseteq$ the following 
Corollary holds.
\begin{corollary}\label{termination_lemma}
There are no infinite sequences of interpretations $I_1 I_2 \ldots I_k\ldots$
such that for all $k$ and for all $j<k$, $I_k\not\sqsubseteq I_j$.
\end{corollary}
Corollary \ref{termination_lemma} ensures that it is not possible 
to generate infinite sequences of interpretations such that each element 
is not {\em subsumed} (using a terminology from constraint logic programming) 
by one of the previous elements in the sequence. 
The problem now is to define a fixpoint operator over 
abstract Herbrand interpretations that is {\em correct} and {\em complete} w.r.t. the ground semantics. 
If we find it, then we can use the corollary to prove that (for any program)
its fixpoint can be reached after finitely many steps. 
For this purpose and using the multiset operations $\setminus$ (difference), 
$\mlub{}{}$ (least upper bound w.r.t. $\preccurlyeq$), 
and $\eps$ (empty multiset) defined in Section \ref{prelim}, 
we first define a new version of the satisfiability relation 
$\models$.
The intuition under the new judgment 
$I\asat \Delta[\calA]$
is that $\calA$ is the {\em minimal} fact (w.r.t. multiset inclusion) 
that should be added to $\Delta$
in order for $\calA+\Delta$ to be satisfiable in $I$.
\begin{definition}[Satisfiability]
\label{spsat}
Let  $I\in\calP(B_P)$, then $\asat$ is defined as follows:
$$\begin{array}{l}
I\asat \all,{\Delta}[{\eps}];\\
[\medskipamount]
I\asat {\calA}[{\calB}\setminus{\calA}]\ \hbox{for}\ {\calB}\in I;\\
[\medskipamount]
I\asat G_1\para G_2,\Delta[{\calA}]\ \hbox{if}\ 
I\asat G_1,G_2,\Delta[{\cal A}];\\
[\medskipamount]
I\asat G_1\with G_2,\Delta[\mlub{\calA_1}{\calA_2}]\ \hbox{if}\ 
I\asat G_1,\Delta[{\calA_1}],\ I\asat G_2,\Delta[{\calA_2}].\\
\end{array}
$$
\end{definition}
Given a finite interpretation $I$ and a context $\Delta$, 
 the previous definition gives us an {\em algorithm} to compute 
 all facts $\calA$ such that $I\asat \Delta[\calA]$ holds.
\begin{example}
\label{exjudge}
Let us consider clause 2. of Example \ref{LOex}, namely
  $$ b\lollo(d\para e)\with f, $$
and $I=\{\{c,d\},\{c,f\}\}$.
We want to compute the facts $\calA$ for which $I\asat G[\calA]$, where
$G=(d\para e)\with f$ is the body of the clause. From the second rule defining
the judgment $\asat$,
we have that $I\asat \{d,e\}[\{c\}]$, because $\{c,d\}\in I$ and
$\{c,d\}\setminus\{d,e\}=\{c\}$. Therefore we get $I\asat d\para e[\{c\}]$
using the third rule for $\asat$. Similarly, we have that
$I\asat\{f\}[\{c\}]$, because $\{c,f\}\in I$ and
$\{c,f\}\setminus\{f\}=\{c\}$.
By applying the fourth rule for $\asat$, with $G_1=d\para e$, $G_2=f$,
$\calA_1=\{c\}$, $\calA_2=\{c\}$ and $\Delta=\eps$,
we get $I\asat G[\{c\}]$, in fact $\mlub{\{c\}}{\{c\}}=\{c\}$.
There are other ways to apply the rules for
$\asat$. In fact, we can get $I\asat \{d,e\}[\{c,f\}]$,
because $\{c,f\}\in I$ and $\{c,f\}\setminus\{d,e\}=\{c,f\}$.
Similarly, we can get $I\asat\{f\}[\{c,d\}]$.
By considering all combinations, it turns out that
$I\asat G[\calA]$, for every $\calA\in\{\{c\},\{c,f\},\{c,d\},\{c,d,f\}\}$.
The information conveyed by $\{c,f\},\{c,d\},\{c,d,f\}$ is in some sense
{\em redundant}, as we shall see in the following (see Example \ref{example}).
In other words, it is not always true that the output fact of the
judgment $\asat$
is {\em minimal} (in the previous example only the output $\{c\}$ is minimal).
Nevertheless, the important point to be stressed here is that the set of
possible facts satisfying the judgment, given $I$ and $G$, is {\em finite}.
This will be sufficient to ensure effectiveness of the fixpoint operator.
\end{example}
The relation $\asat$ satisfies the following properties.
\begin{lemma}\label{abstract_sat_lemma}
For every $I,J\in\calP(B_P)$, context $\Delta$, and fact $\calA$,
\begin{enumerate}
\item[i)]
if $I\asat \Delta[\calA]$, then
$\den{I}\models \Delta[\calA']$ for all $\calA'$ s.t. 
$\calA\preccurlyeq\calA'$;
\item[ii)]
if  $\den{I}\models \Delta[\calA']$, then
there exists $\calA$ such that
$I\asat \Delta[\calA]$ and $\calA\preccurlyeq\calA'$;
\item[iii)]
if $I\asat \Delta[\calA]$ and $I\sqsubseteq J$, then there exists $\calA'$ such
that $J\asat \Delta[\calA']$ and $\calA'\preccurlyeq\calA$;
\item[iv)] given a chain of abstract Herbrand interpretations
$I_1\sqsubseteq I_2\sqsubseteq \ldots$,
if $\den{\bigsqcup_{i=1}^{\infty} I_i}\models \Delta[\calA]$
then there exists $k$ s.t $\den{I_k}\models \Delta[\calA]$.
\end{enumerate}
\end{lemma}
\begin{proof}
\begin{description}
\item[$i)$]
By induction on $\Delta$.
\begin{itemize}
\item[-] $I\asat\all,\Delta[\eps]$  and 
$\den{I}\models\all,\Delta[\calA']$ and
$\eps\preccurlyeq \calA'$ for any $\calA'$;
\item[-]
if $I\asat\calA[\calA']$ then 
$\calA'=\calB\setminus\calA$ for $\calB\in I$. 
Since $\calB\preccurlyeq (\calB\setminus\calA)+\calA=\calA'+\calA$,
we have that $(\calB\setminus\calA)+\calA\in\den{I}$, therefore
$\den{I}\models\calA[\calB\setminus\calA]$, so that $\den{I}\models\calA[\calC]$ for
all $C$ s.t. $\calA'=\calB\setminus\calA\preccurlyeq\calC$, because $\den{I}$
is upward closed;
\item[-]
if $I\asat G_1\with G_2,\Delta[\calA]$ then $\calA=\mlub{\calA_1}{\calA_2}$ 
and $I\asat G_1,\Delta[\calA_1]$ and $I\asat G_2,\Delta[\calA_2]$.
By inductive hypothesis, 
$\den{I}\models G_1,\Delta[\calB_1]$ and 
$\den{I}\models G_2,\Delta[\calB_2]$ for any $\calB_1,\calB_2$ s.t.
$\calA_1\preccurlyeq\calB_1$ and $\calA_2\preccurlyeq\calB_2$. 
That is, $\den{I} \models G_i,\Delta[\calC]$ for any 
$\calC\in \den{\mlub{\calA_1}{\calA_2}}$ $i:1,2$. 
It follows that $\den{I} \models G_1\with G_2,\Delta[\calC]$ for all
$\calC\in \den{\mlub{\calA_1}{\calA_2}}$;
\item[-] the $\para$-case follows by a straightforward
application of the inductive hypothesis.
\end{itemize}
\item[$ii)$]
By induction on $\Delta$.
\begin{itemize}
\item[-] The $\all$-case follows by definition;
\item[-]
if $\den{I}\models\calA[\calA']$ then 
$\calA'+\calA\in \den{I}$, i.e.,  there exists $\calB\in I$ 
s.t. $\calB\preccurlyeq \calA'+\calA$. 
Since $\calB\setminus \calA\preccurlyeq (\calA'+\calA)\setminus\calA=\calA'$,
it follows that for $\calC=\calB\setminus \calA$, 
$I\asat\calA[\calC]$ and $\calC\preccurlyeq \calA'$;
\item[-]
if $\den{I} \models G_1\with G_2,\Delta[\calA]$
then $\den{I} \models G_i,\Delta[\calA]$ for $i:1,2$. 
By inductive hypothesis, there exists $\calA_i$ such that 
$\calA_i\preccurlyeq\calA$,
$I\asat G_i,\Delta[\calA_i]$ for $i:1,2$, i.e., 
$I\asat G_1\with G_2,\Delta[\mlub{\calA_1}{\calA_2}]$.
The thesis follows noting that $\mlub{\calA_1}{\calA_2}\preccurlyeq\calA$;
\item[-] the $\para$-case follows by a straightforward application of the
inductive hypothesis.
\end{itemize}
\item[$iii)$]
If $I\asat \Delta[\calA]$, then by $i)$,
 $\den{I}\models\Delta[\calA]$. 
 Since $\den{I}\subseteq \den{J}$ then, by Lemma \ref{satlemma} $ii)$,
 $\den{J}\models\Delta[\calA]$.
 Thus, by $ii)$,
 there exists $\calA'\preccurlyeq\calA$ s.t. 
 $J\asat\Delta[\calA']$.
\item[$iv)$]
By induction on $\Delta$.
\begin{itemize}
  \item[-] If $\Delta=\all,\Delta'$, then, no matter which $k$ you choose,
    $\den{I_k}\models{\all,\Delta'}[{\calA}]$;
  \item[-] if $\Delta$ is a fact, then 
    $\Delta,\calA\in\den{\bigsqcup_{i=1}^{\infty} I_i}$, that is
    there exists $\calB$ s.t. $\calB\in\bigsqcup_{i=1}^{\infty} I_i$ and
    $\calB\preccurlyeq\Delta,\calA$. Therefore there exists $k$ s.t.
    $\calB\in I_k$ and $\calB\preccurlyeq\Delta,\calA$, that is
    $\Delta,\calA\in\den{I_k}$;
  \item[-] if $\Delta=G_1\with G_2,\Delta'$, then by inductive hypothesis
    there exist $k_1$ and $k_2$ s.t. $\val{\den{I_{k_1}}}{G_1,\Delta'}{\calA}$
    and
    $\val{\den{I_{k_2}}}{G_2,\Delta'}{\calA}$.
    Therefore, if $k=max\{k_1,k_2\}$, by
    Lemma \ref{satlemma} $ii)$, we
    have that $\val{\den{I_k}}{G_1,\Delta'}{\calA}$
    and $\val{\den{I_k}}{G_2,\Delta'}{\calA}$, therefore
    $\val{\den{I_k}}{G_1\with G_2,\Delta'}{\calA}$, i.e.
    $\val{\den{I_k}}{\Delta}{\calA}$;
  \item[-] the $\para$-case follows by a straightforward
    application of the inductive hypothesis.
\end{itemize}
\end{description}
\end{proof}
We are ready now to define the abstract fixpoint operator 
$\STP:\calI\rightarrow\calI$.
We will proceed in two steps. 
We will first define an operator working over 
elements of $\calP(B_P)$.
With a little bit of overloading, we will call the operator 
with the same name, i.e.,  $S_P$.
As for the $\STP$ operator used in the symbolic semantics of CLP programs 
\cite{JM94}, the operator should satisfy the equation  
$\den{\STP(I)}=T_P(\den{I})$ for any $I,~J\in\calP(B_P)$.
This property ensures the soundness and completeness of the 
{\em symbolic} representation w.r.t. the ground semantics of LO programs.

After defining the operator over $\calP(B_P)$, 
we will lift it to our abstract domain $\calI$ consisting 
of the equivalence classes of elements of $\calP(B_P)$ w.r.t. 
the relation $\simeq$ defined in Definition \ref{interpretation_lattice}.
Formally, we first introduce the following definition.
\begin{definition}[Symbolic Fixpoint Operator]\label{fixpoint_operator}
Given an {LO} program $P$, and $I\in\calP(B_P)$,
the operator $\STP$ is defined as follows:
$$
\STP(I)=\{\ms{H}+{\cal A}\ |\ H\leftlolli G\in P,
\ I\asat G[{\cal A}]\}.
$$
\end{definition}
The following property shows that $\STP$ is 
sound and complete w.r.t. $T_P$.
\begin{proposition}\label{completeness} 
Let $I\in\calP(B_P)$, then $\den{\STP(I)}=T_P(\den{I})$.
\end{proposition}
\begin{proof}
Let $\calA=\ms{H},\calB\in \STP(I)$ where $H\leftlolli G\in P$
and $I\asat G[\calB]$ then, by Lemma \ref{abstract_sat_lemma} $i)$,
$\den{I}\models G[\calB']$ for any $\calB'$ s.t. $\calB\preccurlyeq\calB'$.
Thus, for any $\calA'=\ms{H},\calB'$ s.t. $\calA\preccurlyeq\calA'$, 
$\calA'\in T_P(\den{I})$.\\
Vice versa,  if $\calA\in T_P(\den{I})$ then 
$\calA=\ms{H},\calB$ where $H\leftlolli G\in P$
and $\den{I}\models G[\calB]$. 
By Lemma \ref{abstract_sat_lemma} $ii)$, 
there exists $\calB'$ s.t. $\calB'\preccurlyeq\calB$ and 
$I\asat G[\calB']$, i.e., $\calA'=\ms{H},\calB'\in \STP(I)$ 
and $\calA'\preccurlyeq\calA$.
\end{proof}
Furthermore, the following corollary holds.
\begin{corollary}\label{cor_well_def}
Given $I,J\in\calP(B_P)$, if $I\simeq J$ then $S_P(I)\simeq S_P(J)$.
\end{corollary}
\begin{proof}
If $I\simeq J$, then, by definition of $\simeq$, it follows that 
$\den{I}=\den{J}$. This implies that $T_P(\den{I})=T_P(\den{J})$.
Thus, by Prop. \ref{completeness} it follows that 
$\den{S_P(I)}=\den{S_P(J)}$, i.e., $S_P(I)\simeq S_P(J)$.
\end{proof}
The previous corollary allows us to safely lift the definition of $S_P$ 
from the lattice $\tuple{\calP(B_P),\subseteq}$ to the lattice 
$\tuple{\calI,\sqsubseteq}$.
Formally, we define the abstract fixpoint operator as follows. 
\begin{definition}[Abstract Fixpoint Operator $\STP$]\label{symbolic_fixpoint}
Given an {LO} program $P$, and an equivalence class $[I]_\simeq$ of $\calI$, 
the operator $\STP$ is defined as follows:
$$
\STP([I]_\simeq)=[\STP(I)]_\simeq
$$
where $\STP(I)$ is defined in Definition \ref{fixpoint_operator}.
\end{definition}
In the following  we will use $I$ to denote its class $[I]_\simeq$. 
The abstract operator $\STP$ satisfies the following property. 
\begin{proposition}
$\STP$ is monotonic and continuous over the lattice
$\tuple{\calI,\sqsubseteq}$.
\end{proposition}
\begin{proof}
\mbox{}\\
{\em Monotonicity}.
For any $\calA=\ms{H},\calB\in \STP(I)$
there exists $H\leftlolli G\in P$ s.t. 
$I\asat G[\calB]$.
Assume now that $I\sqsubseteq J$. 
Then, by Lemma \ref{abstract_sat_lemma} $iii)$, 
we have that $J\asat G[\calB']$
for $\calB'\preccurlyeq\calB$. Thus, there exists 
$\calA'=\ms{H},\calB'\in \STP(J)$ such that 
$\calA'\preccurlyeq\calA$, i.e., $\STP(I)\sqsubseteq \STP(J)$.
\\
{\em Continuity}. We show that $\STP$ is finitary. 
Let $I_1\sqsubseteq I_2\sqsubseteq \ldots$ be an increasing sequence of 
interpretations.
For any $\calA=\ms{H},\calB\in \STP(\bigsqcup_{i=1}^{\infty} I_i)$ 
there exists $H\leftlolli G\in P$ s.t. 
$\bigsqcup_{i=1}^{\infty} I_i\asat G[\calB]$.
By Lemma \ref{abstract_sat_lemma} $i)$,
$\den{\bigsqcup_{i=1}^{\infty} I_i}\models G[\calB]$. 
By Lemma \ref{abstract_sat_lemma} $iv)$, we
get that $\den{I_k}\models G[\calB]$ for some $k$,
and by Lemma \ref{abstract_sat_lemma} $ii)$, 
$I_k\asat G[\calB']$ for $\calB'\preccurlyeq\calB$.
Thus,  $\calA'=\ms{H},\calB'\in \STP(I_k)$, i.e., 
$\calA'\in \bigsqcup_{i=1}^{\infty} \STP(I_i)$, i.e., 
$\STP(\bigsqcup_{i=1}^{\infty} I_i)\sqsubseteq 
\bigsqcup_{i=1}^{\infty} \STP(I_i)$.
\end{proof}
\begin{corollary}
\label{fixpoint_equivalence}
$\den{\lfp(\STP)}=\lfp(T_P)$.
\end{corollary}
Let $SymbF(P)=\lfp(\STP)$, then we have the following main theorem.
\begin{theorem}[Soundness and Completeness]
\label{main_theorem}
Given an LO program $P$, $O(P)=F(P)=\den{SymbF(P)}$.
Furthermore, there exists $k\in\Nat$ such that 
$SymbF(P)=\bigsqcup_{i=0}^k \isp{k}(\emptyset)$. 
\end{theorem}
\begin{proof}
Theorem \ref{fixpoint_operational} and Corollary \ref{fixpoint_equivalence} 
show that $O(P)=F(P)=\den{SymbF(P)}$. Corollary \ref{termination_lemma} guarantees
that the fixpoint of $\STP$ can always be reached after finitely many steps.
\end{proof}
The previous results give us an algorithm to compute the operational and 
fixpoint semantics of a propositional {LO} program via the operator $\STP$.
The algorithm is inspired by the {\em backward reachability} algorithm 
used in \cite{ACJT96,FS01} to compute {\em backwards} the closure of the 
{\em predecessor} operator of a well-structured transition system.
The algorithm in pseudo-code for computing $F(P)$ is shown in Figure \ref{symbef}.
Corollary \ref{termination_lemma} guarantees that the algorithm always 
terminates and returns a {\em symbolic} representation of $O(P)$. 
\begin{figure*}
\centering
\centerline{
\small
$
\begin{array}{ll}
{\bf Procedure}\ symbF(P:\hbox{{LO}\ program}){\bf :}\\
\ \ \ {\bf set}\ {New}:=\{\ms{H}\ |\ H\leftlolli \all\in P\}\ \ {\bf and}\ \ \hbox{Old}:=\emptyset;\\
\ \ \ {\bf repeat}\\
\ \ \ \ \ \ \ \hbox{Old}:= {Old}\ \cup\  \hbox{New};\\
\ \ \ \ \ \ \ \hbox{New}:=\STP(\hbox{New});\\
\ \ \ {\bf until}\  \hbox{New}\sqsubseteq  \hbox{Old};\\
\ \ \ {\bf return}\ \  \hbox{Old}.
\end{array}
$}
\caption{Symbolic fixpoint computation}
\label{symbef}
\end{figure*} 
As a corollary of Theorem \ref{main_theorem}, we obtain the following result.
\begin{corollary}
The provability of $P\Rightarrow G$ in propositional {LO} is {\em decidable}.
\end{corollary}
In view of Proposition \ref{monotonicity}, this result can be considered as an
instance of the general decidability result \cite{Kop95} for propositional
{\em affine}
linear logic (i.e., linear logic with {\em weakening}).
\begin{example}
\label{example}
We calculate the fixpoint semantics for the program $P$ of Example \ref{LOex},
which is given below.
$$\begin{array}{l}
  1.\ \ a\lollo b\para c\\
  2.\ \ b\lollo(d\para e)\with f\\
  3.\ \ c\para d\lollo\all\\
  4.\ \ e\para e\lollo b\para c\\
  5.\ \ c\para f\lollo\all\\
\end{array}$$
We start the computation from $\isp{0}=\eset$.
The first step consists in adding the multisets corresponding to program
facts, i.e., clauses 3. and 5., therefore we compute
 $$ \isp{1}=\{\{c,d\},\{c,f\}\}. $$
Now, we can try to apply clauses 1., 2., and 4. to facts in $\isp{1}$.
From the first clause, we have that $\isp{1}\asat\{b,c\}[\{d\}]$ and
$\isp{1}\asat\{b,c\}[\{f\}]$, therefore $\{a,d\}$ and $\{a,f\}$ are elements of
$\isp{2}$. Similarly, for clause 2. we have that $\isp{1}\asat\{d,e\}[\{c\}]$
and $\isp{1}\asat\{f\}[\{c\}]$, therefore we have,
from the rule for $\with$, that $\{b,c\}$ belongs to $\isp{2}$ (we can also
derive other judgments for clause 2., as seen in Example \ref{exjudge},
for instance
$\isp{1}\asat\{d,e\}[\{c,f\}]$,
but it immediately turns out that
all these judgments give rise to {\em redundant} information, i.e., facts
that are subsumed by the already calculated ones).
By clause 4., finally we have that $\isp{1}\asat\{b,c\}[\{d\}]$ and
$\isp{1}\asat\{b,c\}[\{f\}]$, therefore $\{d,e,e\}$ and $\{e,e,f\}$ belong to
$\isp{2}$. We can therefore take the
following equivalence class as representative for $\isp{2}$:
$$ \isp{2}=\{\{c,d\},\{c,f\},\{a,d\},\{a,f\},\{b,c\},\{d,e,e\},\{e,e,f\}\}. $$
We can similarly calculate $\isp{3}$. For clause 1. we immediately have
that $\isp{2}\asat\{b,c\}[\eps]$, so that $\{a\}$ is an element of
$\isp{3}$; this makes the information given by $\{a,d\}$ and
$\{a,f\}$ in $\isp{2}$
redundant. From clause 4. we can get that $\{e,e\}$ is another element of
$\isp{3}$, which implies that the information given by $\{d,e,e\}$ and
$\{e,e,f\}$ is now redundant. No additional (not redundant) elements are
obtained from clause 2.
We therefore can take
$$ \isp{3}=\{\{c,d\},\{c,f\},\{b,c\},\{a\},\{e,e\}\}. $$
The reader can verify that
$ \isp{4}=\isp{3}=SymbF(P) $
so that
$$ O(P)=F(P)=\den{\{\{c,d\},\{c,f\},\{b,c\},\{a\},\{e,e\}\}}. $$
We suggest the reader to compare the top-down proof for the goal $e,e$, given
in Figure \ref{LOproof}, and the part of the
bottom-up computation which yields the same
goal. The order in which the backchaining steps are performed is reversed,
as expected. Moreover, the top-down computation requires to solve one
goal, namely $d,e,c$, which is not {\em minimal}, in the sense that its proper
subset $c,d$ is still provable. Using the bottom-up algorithm sketched above,
at every step only the minimal information (in this case $c,d$) is kept at
every step. In general, this strategy has the further advantage of reducing
the amount of non-determinism in the proof search. For instance, let us
consider the goal $b,e,e$ (which is certainly provable because of Proposition
\ref{monotonicity}). This goal has at least two different proofs. The first is
a slight modification of the proof in Figure \ref{LOproof} (just add the
atom $b$ to every sequent). An alternative proof is the following,
obtained by changing the order of applications of the backchaining steps:

$$
    \infer[bc^{(2)}]{P\Rightarrow b,e,e}
   {\infer[\with_r]{P\Rightarrow (d\para e)\with f,e,e}
   {\infer[\para_r]{P\Rightarrow d\para e,e,e}
   {\infer[bc^{(4)}]{P\Rightarrow d,e,e,e}
   {\infer[\para_r]{P\Rightarrow d,e,b\para c}
   {\infer[bc^{(3)}]{P\Rightarrow d,e,b,c}
   {\infer[\all_r]{P\Rightarrow e,b,\all}
   {}}}}}
     &
    \infer[bc^{(4)}]{P\Rightarrow f,e,e}
   {\infer[\para_r]{P\Rightarrow f,b\para c}
   {\infer[bc^{(5)}]{P\Rightarrow f,b,c}
   {\infer[\all_r]{P\Rightarrow b,\all}{}}}}
   }}
$$
There are even more complicated proofs (for instance in the left branch I could
rewrite $b$ again by backchaining over clause 2. rather than axiom 3).
The bottom-up computation avoids these complications by keeping only
{\em minimal} information at every step. We would also like to stress that
the bottom-up computation is always guaranteed to terminate, as stated in
Theorem \ref{main_theorem}, while in general the top-down computation can
diverge.
\end{example}

\section{A Bottom-up Semantics for \LOonet}
\label{one}\label{groundo}
As shown in \cite{And92}, the original formulation of the 
language {LO} can be extended in order to take into consideration 
more powerful programming constructs. 
In this paper we will consider an extension of {LO} where goal formulas 
range over the $\formula{G}$-formulas of Section \ref{lo} and over the logical 
constant $\one$. In other words, we extend {LO} with clauses of the following form:
$$A_1\para\ldots\para A_n\leftlolli \one.$$
We name this language \LOonews,
and use the notation $\dedLOo{P}{\Delta}$ for \LOone sequents.
The meaning of the new kind of clauses is given by the 
following inference scheme:
$$
 \infer[bc\ (H\leftlolli \one\ \in\ P)]
{P\Rightarrow_\one \ms{H}}
{\infer[\one_r]{P\Rightarrow_\one \one}{}}
$$
Note that there cannot be other {\em resources} in the right-hand side 
of the lower sequent apart from $a_1,\ldots,a_n$.
Thus, in contrast with $\all$, the constant $\one$ intuitively
introduces the possibility of 
{\em counting} resources.
\begin{example}
\label{espetri}
LO programs can be used to encode Petri Nets
(see also the proof of Proposition \ref{undecidable} and Section \ref{petri}).
Let us consider a simple Petri net with three places $a$, $b$ and $c$.
We can represent a marking with a multiset of atoms and a transition with a
clause. For instance,
the clause $a~\para~b~\lollo~c~\para~c$
can be interpreted as the Petri Net transition that removes one token from 
place $a$, one token from place $b$, and adds two tokens to place $c$.
By using the constant $\one$, we can specify an operation $trans$
which transfers
{\em all} tokens in place $a$ to place $b$. The encoding is as follows:
$$\begin{array}{l}
  1.\ \ a\para trans\lollo b\para trans\\
  2.\ \ trans\lollo done\with check\\
  3.\ \ check\para b\lollo check\\
  4.\ \ check\para c\lollo check\\
  5.\ \ check\lollo\one\\
\end{array}$$
The first clause specifies the transfer of a single token from $a$ to $b$,
and it is supposed to be used as many times as the number of initial tokens in
$a$. The second clause starts an auxiliary branch of the computation which
checks that all tokens have been moved to $b$.
The proof for the initial marking $a,a,c$ is given in Figure \ref{espetrifig}
(where, for simplicity, applications of the $\para_r$ and $\with_r$ rules
have been
incorporated into backchaining steps).
Note that the check cannot succeed if there are any tokens left in $a$. Using
$\one$ in clause 5. is crucial to achieve this goal.
\begin{figure*}
$$
   \infer[bc^{(1)}]{P\Rightarrow_\one a,a,c,trans}
  {\infer[bc^{(1)}]{P\Rightarrow_\one b,a,c,trans}
  {\infer[bc^{(2)}]{P\Rightarrow_\one b,b,c,trans}
  {\infer[]{P\Rightarrow_\one b,b,c,done}{\vdots}
    &
   \infer[bc^{(3)}]{P\Rightarrow_\one b,b,c,check}
  {\infer[bc^{(3)}]{P\Rightarrow_\one b,c,check}
  {\infer[bc^{(4)}]{P\Rightarrow_\one c,check}
  {\infer[bc^{(5)}]{P\Rightarrow_\one check}
  {\infer[\one_r]{P\Rightarrow_\one \one}{}
  }}}}}}}
$$
\caption{An \LOone proof for the goal $a,a,c,trans$ in the program of
Example \ref{espetri}}
\label{espetrifig}
\end{figure*}
\end{example}
Provability in \LOone amounts to provability in the proof system for {LO} 
augmented with the $\one_r$ rule.
As for {LO}, let us define the top-down operational semantics of an 
\LOone program as follows:

  $$ \opo = \{\calA\ |\ \calA\ is\ a\ fact\ and\ \dedLOo{P}{\calA}\ is\ provable\}. $$
We first note that, in contrast with Proposition \ref{monotonicity}, the weakening rule 
is not admissible in \LOonews. 
This implies that we cannot use the same techniques we used 
for the fragment without $\one$.
So the question is: can we still find a finite representation of $\opo$? 
The following proposition gives us a negative answer.
\begin{proposition}
\label{undecidable}
Given an \LOone program $P$, there is no algorithm to compute $\opo$.
\end{proposition}
\begin{proof}
To prove the result we present an encoding of Vector Addition Systems (VAS) 
as \LOone programs. 
A VAS consists of a transition system defined over $n$ variables
$\tuple{x_1,\ldots,x_n}$ ranging over positive integers.
The transition rules have the form 
$x_1'=x_1+\delta_1,\ldots,x_n'=x_n+\delta_n$ where 
$\delta_n$ is an integer constant.
Whenever $\delta_i<0$,  guards of the form $x_i\geq -\delta_i$ 
ensure that the variables assume only positive values.
Following \cite{Cer95}, the encoding of a VAS in \LOone is defined as follows.
We associate a propositional symbol $a_i\in\Sigma$ 
to each variable $x_i$. A VAS-transition now becomes a 
rewriting rule $H\leftlolli B$ where 
$Occ_{\ms{B}}(a_i)=-\delta_i$ if $\delta_i<0$ (tokens removed from place $i$) 
and $Occ_{\ms{H}}(a_i)=\delta_i$  if $\delta_i\geq 0$ 
(tokens added to place $i$).
We encode the set of initial markings
(i.e., assignments for the variables $x_i$'s) $M_1,\ldots,M_k$ using $k$ clauses
as follows. 
The i-$th$ clause  $H_i\leftlolli\one$ is such that 
if $M_i$ is the assignment $\tuple{x_1=c_1,\ldots,x_n=c_n}$ then 
$Occ_{\ms{H_i}}(a_j)=c_j$ for $j:1,\ldots,n$.
Based on this idea, if $P_V$ is the program that encodes the VAS $V$ 
it is easy to check that  $O(P_V)$ corresponds to 
the set of {\em reachable} markings of $V$ (i.e., to the closure 
$post^*$ of the $successor$ operator $post$ w.r.t. $V$ and the initial
markings). From classical results on Petri Nets (see e.g. the survey \cite{EN94}), 
there is no algorithm to compute the set of reachable states 
of a VAS $V$ (=$O(P_V)$). 
If not so, we would be able to solve the {\em marking equivalence} 
problem that is known to be undecidable.
\end{proof}
Despite Proposition \ref{undecidable}, it is still possible to define a
{\em symbolic}, effective fixpoint operator for \LOone programs
as we will show in the following section.
Before going into more details,  we first rephrase the semantics 
of Section \ref{ground} for \LOonews.
We omit the proofs, which are analogous to those of Section \ref{ground}.
For the sake of
simplicity, in the rest of the paper we will still
denote the satisfiability judgments for \LOone with $\osat$ and $\asat$.
\begin{definition}[Satisfiability in \LOonews]
Let $I$ be a Herbrand interpretation, then $\osat$ is defined as follows:
$$\begin{array}{l} 
  \valo{I}{\one}{\eps};\\
  [\medskipamount]
  \valo{I}{\all,\Delta}{\calA'}\ \hbox{for any fact}\ \calA';\\
  [\medskipamount]
  \valo{I}\calA{\calA'}\ \hbox{if}\ \calA+\calA' \in I;\\
  [\medskipamount]
  \valo{I}{G_1\para G_2,\Delta}{\calA}\ \hbox{if}\ \valo{I}{G_1,G_2,\Delta}{\calA};\\
  [\medskipamount]
  \valo{I}{G_1\with G_2,\Delta}{\calA}\ \hbox{if}\ \valo{I}{G_1,\Delta}{\calA}\ 
    \mbox{and}\ \valo{I}{G_2,\Delta}{\calA}.\\
\end{array}$$
\end{definition}
The new satisfiability relation satisfies the following properties.
\begin{lemma}
\label{movlemmao}\label{satlemmao}\label{finitary_lemmao}
For any interpretations $I,J$,
context $\Delta$, and fact $\calA$,
\begin{enumerate}
\item[i)] $\valo{I}{\Delta}{\calA}\ if\ and\ only\ if\ \valo{I}{\Delta,\calA}{\eps}$;
\item[ii)] if $I\subseteq J$ and $\valo{I}{\Delta}{\calA}$ then $\valo{J}{\Delta}{\calA}$;
\item[iii)]
given a chain of interpretations $I_1\subseteq I_2\subseteq\ldots$, if  
$\valo{\Un{i}{\infty}I_i}{\Delta}{\calA}$ 
then there exists $k$ s.t. $\valo{I_k}{\Delta}{\calA}$.
\end{enumerate}
\end{lemma}
The fixpoint operator $\tpo$ is defined like $T_P$.
\begin{definition}[Fixpoint operator $\tpo$]
Given an {\LOone} program $P$, and an interpretation $I$,
the operator $\tpo$ is defined as follows:
  $$ \tpo(I)=\{\ms{H}+\calA\ |\ H\lollo G\in P,\ \valo{I}{G}{\calA}\}. $$
\end{definition}
The following property holds.
\begin{proposition}\label{montheoro}
$\tpo$ is monotonic and continuous over the lattice
$\tuple{\calD,\subseteq}$.
\end{proposition}
The fixpoint semantics is defined as $\fpo=\lfp(\tpo)=\itpo{\omega}$.
It is sound and complete with respect to the operational semantics, 
as stated in the following theorem.
\begin{theorem}[Soundness and Completeness]
\label{fixpoint_operationalo}
For every \LOone program $P$, $\fpo=\opo$.
\end{theorem}
\section{Constraint Semantics for \LOonet}
\label{nongroundo}
In this section we will define a {\em symbolic} fixpoint operator which relies
on a con\-straint-based representation of provable multisets. 
The application of this operator is effective. 
Proposition \ref{undecidable} shows however that there is no guarantee that its fixpoint 
can be reached after finitely many steps.
According to the encoding of VAS used in the proof of Proposition \ref{undecidable},
let $\vec{x}=\tuple{x_1,\ldots,x_n}$ be a vector of variables, where variable $x_i$ 
denotes the number of
occurrences of $a_i\in\Sigma$ in a given fact. Then we can
immediately recover the semantics of Section \ref{nonground} using a
very simple class of linear constraints over integer variables.
Namely, given a fact $\calA$
we can denote its closure, i.e., the ideal $\den{\calA}$, by the constraint
$$\varphi_{\den{\cal A}}\ \equiv\ \bigwedge_{i=1}^{n}x_i\geq Occ_{\cal A}(a_i).$$
Then all the operations on multisets involved in the definition of
$\STP$ (see Definition \ref{spsat}) can be expressed as operations over
linear constraints. In particular, given the ideals
$\den{\cal A}$ and $\den{\cal B}$, the ideal $\den{\calA\bullet\calB}$
is represented as the constraint
  $$ \varphi_{\den{{\cal A}\bullet{\cal B}}}=
     \varphi_{\den{\cal A}}\wedge\varphi_{\den{\cal B}}, $$ while
$\den{\calB\setminus\calA}$, for a given multiset $\calA$, is represented as
the constraint
  $$ \varphi_{\den{{\cal B}\setminus{\cal A}}}=
     \exists\vec{x'}.(\varphi_{\den{\cal B}}[\vec{x}'/\vec{x}]\wedge 
      \Rem_{\cal A}(\vec{x},\vec{x'})), $$
     where  
  $$\Rem_{\cal A}(\vec{x},\vec{x'})\ \equiv\ 
   \bigwedge_{i=1}^n x_i=x_i'-Occ_{\cal A}(a_i)\ \wedge\ x_i\geq 0.$$
The constraint $\Rem_{\cal A}$ models the removal of the occurrences of the literals 
in ${\calA}$ from all elements of the denotation of ${\calB}$.
Similarly, $\den{\calB+\calA}$, for a given multiset $\calA$, 
is represented as the constraint 
  $$ \varphi_{\den{{\cal B}+{\cal A}}}=\exists \vec{x'}.(\varphi_{\den{\cal B}}[\vec{x}'/\vec{x}]\wedge \Add_{\cal A}(\vec{x},\vec{x'})), $$
where 
$$\Add_{\cal A}(\vec{x},\vec{x'})\ 
\equiv\ \bigwedge_{i=1}^n 
x_i=x_i'+Occ_{\cal A}(a_i).$$
The introduction of the constant $\one$ breaks down Proposition \ref{monotonicity}. 
As a consequence, the abstraction based on ideals is no more precise.
In order to give a semantics for \LOonews, we need to add a class of
constraints for representing collections of multisets which are not 
upward-closed (i.e., which are not ideals). 
We note then that we can represent a multiset ${\cal A}$ as the linear constraint
$$\varphi_{\cal A}\ \equiv\ \bigwedge_{i=1}^{n}x_i=Occ_{\cal A}(a_i).$$
The operations over linear constraints discussed previously extend smoothly
when adding this new class of equality constraints. In particular,
given two constraints $\varphi$ and $\psi$,
their conjunction $\varphi\cand\psi$ still plays the role that the operation
$\bullet$ (least upper bound of multisets) had in Definition \ref{spsat}, while
$\exists\vec{x'}.(\varphi[\vec{x}'/\vec{x}]\wedge \Rem_{\cal A}(\vec{x},\vec{x'}))$, for a given multiset $\calA$, plays the role of multiset difference.
The reader can compare Definition \ref{spsat} with Definition \ref{sposat}. 
Based on these ideas, we can define a bottom-up evaluation procedure for 
\LOone programs via an extension $\STPo$ of the operator $\STP$.

In the following we will use the notation $\msc{\vec{c}}$, where
$\vec{c}=\tuple{c_1,\ldots,c_n}$ is
a solution of a constraint $\varphi$ (i.e., an assignment of natural numbers
to the variables $\vec{x}$ which satisfies $\varphi$),
to indicate the multiset over $\Sigma=\{a_1,\ldots,a_n\}$
which contains $c_i$ occurrences of every propositional symbol $a_i$
(i.e., according to the notation introduced above,
$\vec{c}$ is the unique solution of
$\varphi_{\msc{\vec{c}}}$).
We extend this definition to a set $C$ of constraint
solutions by $\msc{C}=\{\msc{\vec{c}}\ |\ \vec{c}\in C\}$.
We then define the denotation of a given constraint
$\varphi$, written $\deno{\varphi}$,
as the set of multisets corresponding to solutions of $\varphi$, i.e.,
$\deno{\varphi}=\{\msc{\vec{c}}\ |\ \vec{x}=\vec{c}\ \hbox{satisfies}\ \varphi\}$.

Following \cite{GDL95}, we introduce an equivalence relation $\simeq$ over constraints, given by
$\varphi\simeq\psi$ if and only if $\deno{\varphi}=\deno{\psi}$, i.e.,
we identify constraints with the same set of solutions. For the
sake of simplicity, in the following we will identify a constraint with its
equivalence class, i.e., we will simply write $\varphi$ instead of
$[\varphi]_\simeq$. 
Let $\linc$ be the set of (equivalence classes of) of linear constraints
over the integer variables 
$\vec{x}=\tuple{x_1,\ldots,x_n}$ 
associated to the signature $\Sigma=\{a_1,\ldots,a_n\}$.
The operator $\STPo$ is defined on {\em constraint interpretations} consisting 
of sets (disjunctions) of (equivalence classes of) linear constraints.
For brevity, we will define the semantics directly on the interpretations
consisting of the representative elements of the equivalence classes. 
The denotation $\deno{I}$ of a constraint interpretation $I$ extends the one for
constraints as expected:
$\deno{I}=\{\deno{\varphi}\ |\ \varphi\in I\}$.
Interpretations form a complete lattice with respect to
set inclusion.
\begin{definition}[Constraint Interpretation]
We say that $I\subseteq\linc$ is a constraint interpretation. Constraint
interpretations form a complete lattice $\tuple{\calC,\subseteq}$
with respect to set inclusion, where $\calC=\calP(\linc)$.
\end{definition}
We obtain then a new notion of satisfiability using operations over constraints as follows.
In the following definitions we assume that the conditions apply only when 
the constraints are {\bf satisfiable} (e.g. $x=0\wedge x\geq 1$ 
has no solutions thus the following rules cannot be applied to this case).
\begin{definition}[Satisfiability in \LOonews]
\label{sposat}
Let $I\in\calC$, then $\asat$ is defined as follows: 
$$\begin{array}{l}
\symbvalo{I}{\one}{\varphi}
\ \hbox{where}\ \varphi\ \equiv\  x_1=0\wedge\ldots\wedge x_n=0;\\
[\medskipamount]
\symbvalo{I}{\all,{\Delta}}{\varphi}\ \hbox{where}\ 
\varphi\ \equiv\  x_1\geq 0\wedge \ldots\wedge x_n\geq 0;\\
[\medskipamount]
\symbvalo{I}{\calA}{\varphi}\ \hbox{where}\ 
\varphi\equiv\exists\vec{x}'.(\psi[\vec{x}'/\vec{x}]\wedge \Rem_{\cal A}(\vec{x},\vec{x'})),
\ \psi\in I;\\
[\medskipamount]
\symbvalo{I}{G_1\para G_2,\Delta}{\varphi}\ \hbox{if}\  
\symbvalo{I}{G_1,G_2,\Delta}{\varphi};\\
[\medskipamount]
\symbvalo{I}{G_1\with G_2,\Delta}{\varphi_1\wedge\varphi_2}\ \hbox{if}\ 
\symbvalo{I}{G_1,\Delta}{\varphi_1},\ 
\symbvalo{I}{G_2,\Delta}{\varphi_2}.
\end{array}
$$
\end{definition}
The relation $\asat$ satisfies the following properties.
\begin{lemma}
\label{abstract_sat_one_properties}
Given $I,J\in\calC$, 
\begin{enumerate}
\item[i)] if $\symbvalo{I}{\Delta}{\varphi}$, then
         $\valo{\deno{I}}{\Delta}{\calA}$ for every $\calA\in\deno{\varphi}$;
\item[ii)] if $\valo{\deno{I}}{\Delta}{\calA}$, then
there exists $\varphi$ such that $\symbvalo{I}{\Delta}{\varphi}$ and
$\calA\in\deno{\varphi}$;
\item[iii)] if $I\subseteq J$ and $\symbvalo{I}{\Delta}{\varphi}$, 
then $\symbvalo{J}{\Delta}{\varphi}$;
\item[iv)]
given a chain of constraint interpretations 
$I_1\subseteq I_2\subseteq\ldots$, 
if $\bigcup_{i=1}^\infty \symbvalo{I_i}{\Delta}{\varphi}$ 
then there exists $k$ s.t. $\symbvalo{I_k}{\Delta}{\varphi}$.
\end{enumerate}
\end{lemma}
\begin{proof}
\begin{description}
\item[$i)$]
By induction on $\Delta$.
\begin{itemize}
\item[-] If $I\asato\all,\Delta[\varphi]$, then every $\vec{c}$ (with $c_i\geq 0$)
  is solution of $\varphi$, and $\valo{\deno{I}}{\all,\Delta}{\calA'}$
  for every fact $\calA'$;
\item[-] if $I\asato\one[\varphi]$, then $\tuple{0,\ldots,0}$ is the only
 solution of $\varphi$, and $\valo{\deno{I}}{\one}{\eps}$;
\item[-] if $I\asato\calA[\varphi]$ then
  there exists $\psi\in I$ s.t. $\varphi\equiv\exists \vec{x'}.
  (\psi[\vec{x}'/\vec{x}]\wedge \rho_{\cal A}(\vec{x},\vec{x'}))$
  is satisfiable.
  Then for every solution $\vec{c}$ of $\varphi$ there exists a vector
  $\vec{c'}$ s.t.
  $\psi[\vec{c'}/\vec{x}]$ is satisfiable and
  $c_1'\geq Occ_{\cal A}(a_1),c_1=c_1'-Occ_{\cal A}(a_1),\ldots,
  c_n'\geq Occ_{\cal A}(a_n),c_n=c_n'-Occ_{\cal A}(a_n)$. From this we get
  that for $i=1,\ldots,n$, $c_i'=c_i+Occ_{\cal A}(a_i)$ is a solution for
  $\psi$, therefore $\msc{\vec{c}}+\calA\in\deno{\psi}
  \subseteq\deno{I}$ so that we
  can conclude
  $\valo{\deno{I}}{\calA}{\msc{\vec{c}}}$;
\item[-] if $I\asato G_1\with G_2,\Delta[\varphi]$ then
  $\varphi\equiv\varphi_1\cand\varphi_2$ and
  $I\asato G_1,\Delta[\varphi_1]$, $I\asato G_2,\Delta[\varphi_2]$. By
  inductive hypothesis, $\valo{\deno{I}}{G_1,\Delta}{\msc{\vec{c_1}}}$
  and $\valo{\deno{I}}{G_2,\Delta}{\msc{\vec{c_2}}}$ for every
  $\vec{c_1}$ and $\vec{c_2}$ solutions of $\varphi_1$ and $\varphi_2$,
  respectively. Thus
  $\valo{\deno{I}}{G_1\with G_2,\Delta}{\msc{\vec{c}}}$ for every
  $\vec{c}$ which is solution of both $\varphi_1$ and $\varphi_2$, i.e.
  for every $\vec{c}$ which is solution of $\varphi_1\cand\varphi_2$;
\item[-] the $\para$-case follows by a straightforward
  application of the inductive hypothesis.
\end{itemize}
\item[$ii)$]
By induction on $\Delta$.
\begin{itemize}
\item[-] $\valo{\deno{I}}{\all,\Delta}{\msc{\vec{c}}}$ for every $\vec{c}$,
  and $I\asato\all,\Delta[\varphi]$, where
  $\varphi\equiv x_1\geq 0,\ldots,x_n\geq 0$, and every $\vec{c}$ is solution
  of $\varphi$;
\item[-] if $\valo{\deno{I}}{\one}{\eps}$, $\eps=\msc{\tuple{0,\ldots,0}}$,
  then $I\asato\one[\varphi]$, where
  $\varphi\equiv x_1=0,\ldots,x_n=0$, and $\tuple{0,\ldots,0}$ is solution
  of $\varphi$;
\item[-] if $\valo{\deno{I}}{\calA}{\msc{\vec{c}}}$ then
  $\msc{\vec{c}}+\calA\in\deno{I}$, therefore  there exists
  $\psi\in I$ s.t. $\msc{\vec{c}}+\calA\in\deno{\psi}$. Therefore,
  if $\vec{a}$ is such that $\msc{\vec{a}}=\calA$, we have that
  $\psi[\vec{c}+\vec{a}/x]$ is satisfiable, $\vec{c}$ is solution of
  $\varphi\equiv\exists \vec{x'}.
  (\psi[\vec{x}'/\vec{x}]\cand \rho_{\cal A}(\vec{x},\vec{x'}))$
  and $I\asato\calA[\varphi]$;
\item[-] if $\valo{\deno{I}}{G_1\with G_2,\Delta}{\msc{\vec{c}}}$ then
  $\valo{\deno{I}}{G_1,\Delta}{\msc{\vec{c}}}$ and
  $\valo{\deno{I}}{G_2,\Delta}{\msc{\vec{c}}}$. By inductive hypothesis,
  there exist $\varphi_1$ and $\varphi_2$ such that
  $I\asato G_1,\Delta[\varphi_1]$ and $I\asato G_2,\Delta[\varphi_2]$,
  and $\vec{c}$ is a solution of $\varphi_1$ and $\varphi_2$.
  Therefore $I\asato G_1\with G_2,\Delta[\varphi_1\cand\varphi_2]$
  and $\vec{c}$ is a solution of $\varphi_1\cand\varphi_2$;
\item[-] the $\para$-case follows by a straightforward
  application of the inductive hypothesis.
\end{itemize}
\item[$iii)$]
By simple induction on $\Delta$.
\item[$iv)$]
By induction on $\Delta$.
\begin{itemize}
\item[-] The $\all$ and $\one$-cases follow by definition;
\item[-] if $\bigcup_{i=1}^\infty I_i\asato\calA[\varphi]$,
  then there exists $\psi\in\bigcup_{i=1}^\infty I_i$ s.t.
  $\varphi\equiv\exists \vec{x'}.
  (\psi[\vec{x}'/\vec{x}]\cand \rho_{\cal A}(\vec{x},\vec{x'}))$
  is satisfiable. Then there exists $k$ s.t. $\psi\in I_k$
  and $I_k\asato\calA[\varphi]$;
\item[-] if $\bigcup_{i=1}^\infty I_i\asato G_1\with G_2,\Delta[\varphi]$,
  then $\varphi\equiv\varphi_1\cand\varphi_2$, and, by inductive hypothesis,
  there exist $k_1$ and $k_2$ s.t. $I_{k_1}\asato G_1,\Delta[\varphi_1]$ and
  $I_{k_2}\asato G_2,\Delta[\varphi_2]$. Then, for $k=max\{k_1,k_2\}$, we
  have, by $iii)$,
  $I_{k}\asato G_1,\Delta[\varphi_1]$ and $I_{k}\asato G_2,\Delta[\varphi_2]$,
  therefore $I_{k}\asato G_1\with G_2,\Delta[\varphi_1\cand\varphi_2]$;
\item[-] the $\para$-case follows by a straightforward
  application of the inductive hypothesis.
\end{itemize}
\end{description}
\end{proof}
We are now ready to define the extended operator $\STPo$.
\begin{definition}[Symbolic Fixpoint Operator $\STPo$]
\label{symbolic_fixpoint_with_one}
Given an \LOone program $P$, and $I\in\calC$,
the operator $\STPo$ is defined as follows:
$$
\begin{array}{lcl}
\STPo(I)=\{ 
\ \varphi & | 
& H\leftlolli G\in P,\ \symbvalo{I}{G}{\psi},\\
& & \varphi\equiv\exists\vec{x'}.(\psi[\vec{x'}/\vec{x}]\wedge \Add_{\ms{H}}(\vec{x},\vec{x'}))		
\}.
\end{array}
$$ 
\end{definition}
The new operator satisfies the following property.
\begin{proposition}
\label{constraints_monotonicity}
\label{constraints_continuity}
The  operator $\STPo$ is monotonic and continuous over the lattice
$\tuple{\calC,\subseteq}$.
\end{proposition}
\begin{proof}
{\em Monotonicity}. Immediate from $\STPo$ definition and Lemma
\ref{abstract_sat_one_properties} $iii)$.\\
{\em Continuity}.
Let $I_1\subseteq I_2\subseteq\ldots$, be
an increasing sequence of interpretations. 
We show that 
$\STPo(\bigcup_{i=1}^\infty I_i)\subseteq 
\bigcup_{i=1}^\infty \STPo(I_i)$.
If $\varphi\in \STPo(\bigcup_{i=1}^\infty I_i)$, 
by definition there exists a clause $H\leftlolli G\in P$ s.t. 
$\bigcup_{i=1}^\infty I_i\asato G[\psi]$
and $\varphi\equiv\exists\vec{x'}.(\psi[\vec{x'}/\vec{x}]
\wedge \alpha_{\ms{H}}(\vec{x},\vec{x'}))$ is satisfiable.
By Lemma \ref{abstract_sat_one_properties} $iv)$, there exists
$k$ s.t. $I_k\asato G[\psi]$. 
This implies that $\varphi\in\STPo(I_k)$, i.e., 
$\varphi\in\bigcup_{i=1}^\infty \STPo(I_i)$.
\end{proof}
Furthermore, $\STPo$ is a symbolic version of the ground operator $\tpo$,
as stated below.
\begin{proposition}
Let $I\in\calC$, then $\deno{\STPo(I)}=T_P^\one(\deno{I})$.
\end{proposition}
\begin{proof}
Let $\msc{\vec{c}}\in\deno{\STPo(I)}$,
then there exist $\varphi\in\STPo(I)$ and
a clause $H\lollo G\in P$, s.t.
$\vec{c}$ is solution of $\varphi$,
$\varphi\equiv\exists\vec{x'}.
(\psi[\vec{x}'/\vec{x}]\wedge \alpha_{\cal\ms{H}}(\vec{x},\vec{x'}))$ and
$I\asato G[\psi]$.
Then there exists $\vec{c'}$ solution of $\psi$ s.t.
$\msc{\vec{c}}=\msc{\vec{c'}}+\msc{H}$, and, by Lemma
\ref{abstract_sat_one_properties} $i)$, $\valo{\deno{I}}{G}{\msc{\vec{c'}}}$.
Therefore, by definition of $\tpo$,
$\msc{\vec{c}}=\msc{\vec{c'}}+\msc{H}\in\tpo(\deno{I})$.
Vice versa,
let $\msc{\vec{c}}\in\tpo(\deno{I})$,
then there exists $H\lollo G\in P$ s.t. $\valo{\deno{I}}{G}{\calA}$ and 
$\msc{\vec{c}}=\msc{H}+\calA$.
By Lemma \ref{abstract_sat_one_properties} $ii)$, there exists $\psi$ s.t.
$I\asato G[\psi]$ and $\calA\in\deno{\psi}$. Therefore
$\varphi\equiv\exists\vec{x'}.
(\psi[\vec{x'}/\vec{x}]\cand \alpha_{\ms{H}}(\vec{x},\vec{x'}))\in\STPo(I)$,
and $\msc{\vec{c}}=\msc{H}+\calA\in\deno{\varphi}\subseteq\deno{\STPo(I)}$.
\end{proof}
\begin{corollary}
\label{fixpoint_equivalenceo}
$\deno{\lfp(\STPo)}=\lfp(\tpo)$.
\end{corollary}
Now, let $SymbF_\one(P)=\lfp(\STPo)$, then we have the following main theorem
that shows that $\STPo$ can be used (without termination guarantee) to compute symbolically the 
set of logical consequences of an {\LOone} program. 
\begin{theorem}[Soundness and completeness]
Given an \LOone program $P$,
 $\opo=\fpo$ $=\deno{SymbF_\one(P)}$.
\end{theorem}
\begin{proof}
By Theorem \ref{fixpoint_operationalo} and Corollary
\ref{fixpoint_equivalenceo}.
\end{proof}

\section{Bottom-up Evaluation for \LOonet}
\label{nongroundo_computing}
Using a constraint-based representation for \LOone provable multisets,
we have reduced the problem of computing $O_\one(P)$ to the problem 
of computing the reachable states of a system with {\em integer} variables.
As shown by Proposition \ref{undecidable}, the termination of the algorithm 
is not guaranteed a priori. In this respect, Theorem \ref{main_theorem}
gives us sufficient conditions that ensure its termination.
The symbolic fixpoint operator $\STPo$ of Section \ref{nongroundo}
is defined over the lattice $\tuple{\calP(\linc),\subseteq}$, 
with set inclusion being
the partial order relation and set union the least upper bound operator.
When we come to a concrete implementation of $\STPo$, it is worth considering
a weaker ordering relation between interpretations, namely pointwise
{\em subsumption}. Let $\preq$ be the partial order between (equivalence
classes of) constraints given by $\varphi\preq\psi$ if and only if
$\deno{\psi}\subseteq\deno{\varphi}$. Then we say that an interpretation $I$
is subsumed by an interpretation $J$, written $I\sqsubseteq J$, if and only
if for every $\varphi\in I$ there exists $\psi\in J$ such that
$\psi\preq\varphi$.

As we do not need to distinguish between
different interpretations representing the
same set of solutions, we can consider interpretations $I$ and $J$ to be
equivalent
in case both $I\sqsubseteq J$ and $J\sqsubseteq I$ hold.
In this way, we get a lattice of interpretations ordered by $\sqsubseteq$
and such that the least upper bound operator is still set union.
This construction is the natural extension of the one of Section
\ref{nonground}. Actually, when we limit ourselves to considering {LO} programs
(i.e., without the constant $\one$) it turns out that we need only consider
constraints of the form $\vec{x}\geq\vec{c}$, which can be abstracted away
by considering the upward closure of $\msc{\vec{c}}$, as we did in Section
\ref{nonground}. The reader can note that the $\preq$ relation defined above
for constraints is an extension of the multiset inclusion relation we used in
Section \ref{nonground}.

The construction based on $\sqsubseteq$
can be directly incorporated into the semantic
framework presented in Section \ref{nongroundo}, where,
for the sake of simplicity, we have adopted an approach based on $\subseteq$.
Of course, relation $\subseteq$ is stronger than $\sqsubseteq$, therefore
a computation based on $\sqsubseteq$ is correct and it terminates every time
a computation based on $\subseteq$ does.
However, the converse does not always hold, and
this is why a concrete algorithm for computing the least fixpoint of
$\STPo$ relies on subsumption. Let us see an example.

\begin{example}
\label{symbex}
We calculate the fixpoint semantics for the following \LOone program made
up of six clauses:
$$\begin{array}{l}
  1.\ \ a\lollo\one\\
  2.\ \ a\para b\lollo\all\\
  3.\ \ c\para c\lollo\all\\
  4.\ \ b\para b\lollo a\\
  5.\ \ a\lollo b\\
  6.\ \ c\lollo a\with b\\
\end{array}$$
\begin{figure*}
\centering
$$\begin{array}{l}
\begin{array}{lllllll}
  \isp{1} & = & \{ & x_a=1\cand x_b=0\cand x_c=0,             &
                     x_a\geq 1\cand x_b\geq 1\cand x_c\geq 0, \\
          & &      & x_a\geq 0\cand x_b\geq 0\cand x_c\geq 2  & \} 
\end{array}
\\
[\medskipamount]
\begin{array}{lllllll}
  \isp{2} & = & \{ & x_a=1\cand x_b=0\cand x_c=0,             &
                     x_a\geq 1\cand x_b\geq 1\cand x_c\geq 0, \\
          &   &    & x_a\geq 0\cand x_b\geq 0\cand x_c\geq 2, & 
                     x_a=0\cand x_b=2\cand x_c=0,  \\
	  &   &    & x_a\geq 0\cand x_b\geq 3\cand x_c\geq 0, &
                     x_a\geq 2\cand x_b\geq 0\cand x_c\geq 0  & \} 
\end{array}
\\
[\medskipamount]
\begin{array}{lllllll}
  \isp{3} & = & \{ & x_a=1\cand x_b=0\cand x_c=0,             &
                     x_a\geq 1\cand x_b\geq 1\cand x_c\geq 0, \\
          &   &    & x_a\geq 0\cand x_b\geq 0\cand x_c\geq 2, &  
	             x_a=0\cand x_b=2\cand x_c=0,             \\
          &   &    & x_a\geq 0\cand x_b\geq 3\cand x_c\geq 0, &
                     x_a\geq 2\cand x_b\geq 0\cand x_c\geq 0, \\
          &   &    & x_a=0\cand x_b=1\cand x_c=1,             &
                     x_a\geq 0\cand x_b\geq 2\cand x_c\geq 1, \\
          &   &    & x_a\geq 1\cand x_b\geq 0\cand x_c\geq 1  & \} 
\end{array}
\end{array}
$$
\caption{Symbolic fixpoint computation for the program in Example \ref{symbex}}
\label{symbexample}
\end{figure*} 
Let $\Sigma=\{a,b,c\}$ and consider constraints over the variables
$\vec{x}=\tuple{x_a,x_b,x_c}$.
We have that
$\isp{0}=\eset\asat\one[x_a=0\cand x_b=0\cand x_c=0]$, therefore,
by the first clause, $\varphi\in\isp{1}$, where
$\varphi=\exists\vec{x'}.(x'_a=0\cand x'_b=0\cand x'_c=0\cand x_a=x'_a+1\cand
 x_b=x'_b\cand x_c=x'_c)$, which is equivalent to
$x_a=1\cand x_b=0\cand x_c=0$. From now on, we leave to the reader the details
concerning equivalence of constraints.
By reasoning in a similar way, using clauses 2. and 3. we calculate $\isp{1}$
(see Figure \ref{symbexample}).

We now compute $\isp{2}$.
By 4., as $\isp{1}\asat a[x_a=0\cand x_b=0\cand x_c=0]$, we get
$x_a=0\cand x_b=2\cand x_c=0$, and,
similarly, we get $x_a\geq 0\cand x_b\geq 3\cand x_c\geq 0$.
By 5., we have $x_a\geq 2\cand x_b\geq 0\cand x_c\geq 0$,
while clause 6. is not (yet) applicable. Therefore, modulo redundant
constraints (i.e., constraints {\em subsumed} by the already calculated ones)
the value of $\isp{2}$ is given in Figure \ref{symbexample}.

Now, we can compute $\isp{3}$.
By 4. and $x_a\geq 2\cand x_b\geq 0\cand x_c\geq 0\in\isp{2}$ we get
$x_a\geq 1\cand x_b\geq 2\cand x_c\geq 0$, which is subsumed by
$x_a\geq 1\cand x_b\geq 1\cand x_c\geq 0$. By 5. and
$x_a=0\cand x_b=2\cand x_c=0$, we get
$x_a=1\cand x_b=1\cand x_c=0$, subsumed by
$x_a\geq 1\cand x_b\geq 1\cand x_c\geq 0$. Similarly, by 5. and
$x_a\geq 0\cand x_b\geq 3\cand x_c\geq 0$ we get redundant information.
By 6., from
$x_a\geq 1\cand x_b\geq 1\cand x_c\geq 0$ and
$x_a=0\cand x_b=2\cand x_c=0$ we get
$x_a=0\cand x_b=1\cand x_c=1$, from
$x_a\geq 1\cand x_b\geq 1\cand x_c\geq 0$ and
$x_a\geq 0\cand x_b\geq 3\cand x_c\geq 0$ we get
$x_a\geq 0\cand x_b\geq 2\cand x_c\geq 1$, and finally from
$x_a\geq 2\cand x_b\geq 0\cand x_c\geq 0$ and
$x_a\geq 1\cand x_b\geq 1\cand x_c\geq 0$ we have
$x_a\geq 1\cand x_b\geq 0\cand x_c\geq 1$.
The reader can verify that no additional provable multisets can be obtained.
It is somewhat tedious, but in no
way difficult, to verify that clause 6. yields only redundant information
when applied to every possible couple of constraints in $\isp{3}$.
We have then
$\isp{4}=\isp{3}=SymbF_\one(P)$, so that in this particular case we achieve
termination. We can reformulate the operational semantics of $P$ using the
more suggestive multiset notation (we recall that
$\den{\calA}=\{\calB\ |\ \calA\preq\calB\}$, where $\preq$ is multiset
inclusion):
$$\begin{array}{c}
\fpo=\{\{a\},\{b,b\},\{b,c\}\}\ \cup
\den{\{a,b\},\{c,c\},\{b,b,b\},\{a,a\},\{b,b,c\},\{a,c\}}.
\end{array}$$
\end{example}

It is often not the case that the symbolic computation of \LOone
program semantics can be carried out in a finite number of steps. 
Nevertheless,
it is important to remark that viewing the bottom-up evaluation of
\LOone programs
as a least fixpoint computation over infinite-state {\em integer} systems
allows us to apply techniques and tools developed in {\em infinite-state}
model checking 
(see e.g. \cite{ACJT96,BGP97,DP99,FS01,HHW97}) and program analysis \cite{CH78}
to compute approximations of the least fixpoint of $\STPo$.

In the next section we will
present an interesting application of the semantical framework we have
presented so far. Namely, we shall make a detailed comparison between LO
and Disjunctive Logic Programming. This will help us in clarifying the
relations and the relative strength of the languages. After recalling the basic
definitions of DLP in Section \ref{dlplo}, we will present our view of
DLP as an abstraction of LO in Section \ref{comparison}. Finally,
in Section \ref{petri} we will
give a few hints on how to employ this framework to
study reachability problems in Petri Nets.
\section{An Application of the Semantics: Relation with DLP}
\label{dlplo}
As anticipated in the introduction, the paradigms of linear logic programming 
and Disjunctive Logic Programming have in common the use of {\em multi-headed} clauses.
However, the operational interpretation of the extended notion of clause 
is quite different in the two paradigms.
In fact, as shown in \cite{BDM00b}, from a proof-theoretical perspective it is
possible to view LO as a {\em sub-structural} fragment of DLP in which the
 rule of {\em contraction} is forbidden on the right-hand side of sequents. 

While proof theory allows one to compare the {\em top-down}
semantics of the two languages, abstract interpretation \cite{CC77} can be
used to relate the fixpoint, bottom-up evaluation of programs. 
In the following we will focus on the latter approach, exploiting our semantics of LO 
and the bottom-up semantics of DLP given in \cite{MRL91}.
For the sake of clarity, we will use superscripts in order to distinguish between 
the fixpoint operators for LO and DLP, which will be
denoted by $\lotp$ and $\dtp$, respectively. 
First of all, we recall some definitions concerning Disjunctive Logic
Programming.

A {\em disjunctive logic program} as defined in \cite{MRL91} is a 
finite set of clauses  
$$A_1\clor\ldots\clor A_n\cimp B_1\cand\ldots\cand B_m,$$ 
where $n\geq 1$, $m\geq 0$, and $A_i$ and $B_i$ are atomic formulas.
A {\em disjunctive goal} is of the form $\cimp C_1,\ldots,C_n$, where 
$C_i$ is a {\em positive clause} (i.e., a disjunction of atomic formulas) 
for $i:1,\ldots,n$.
To make the language symmetric, in this paper we will consider 
extended clauses of the form
$$A_1\clor\ldots\clor A_n\cimp C_1\cand\ldots\cand C_m$$ 
containing positive clauses in the body.
Following \cite{MRL91},
we will identify positive clauses with sets of atoms.
In order to define the operational and denotational semantics of DLP, 
we need the following notions.
\begin{definition}[Disjunctive Herbrand Base]
The {\em disjunctive Herbrand base} of a program $P$, for short $\dhbp$, 
is the set of 
all positive clauses formed by an arbitrary number of atoms. 
\end{definition}
\begin{definition}[Disjunctive Interpretation]
A subset $I$ of the disjunctive Her\-brand base $\dhbp$ 
is called a disjunctive Herbrand interpretation. 
\end{definition}
\begin{definition} [Ground SLO-derivation]\label{slo-derivation}
Let $P$ be a DLP program. An SLO-derivation of a ground goal $G$ from $P$ 
consists of a sequence of goals $G_0=G,G_1,\ldots$ such that for 
all $i\geq 0$, $G_{i+1}$ is obtained
from $G_i=\leftarrow (C_1,\ldots,C_m,\ldots,C_k)$ as follows:
\begin{itemize}
\item[-] $C\cimp D_1\cand\ldots\cand D_q$ is a ground instance of a clause in $P$ such that
$C$ is contained in $C_m$ (the selected clause);
\item[-]  $G_{i+1}$ is the goal
 $\leftarrow (C_1,\ldots,C_{m-1},D_1\clor C_m,\ldots,D_q\clor C_m,C_{m+1},\ldots,C_k)$.
\end{itemize}
\end{definition}
\begin{definition} [SLO-refutation]
Let $P$ be a DLP program. An SLO-refutation of a ground goal $G$ from $P$ 
is an SLO-derivation $G_0,G_1,\ldots,G_k$ s.t. $G_k$ consists of the empty clause only.
\end{definition}
As SLD-resolution for Horn programs, SLO-resolution gives us a procedural interpretation 
of DLP programs.
The operational semantics is defined then as follows:
$$\dosp=\{ C \ |\ C\in\dhbp,\ \leftarrow C\ \mbox{has\ an\ SLO-refutation}\}.$$
As for Horn programs, it is possible to define a fixpoint semantics via the following operator (where $\gnd(P)$ denotes the set of ground instances of clauses
in $P$).
\begin{definition} [The $\dtp$ Operator]
Given a DLP program $P$ and $I\subseteq \dhbp$,
$$\begin{array}{lll}
\dtp(I) & = & \{\ C\in \dhbp\ |\ C'\cimp D_1,\ldots,D_n\in\gnd(P),\\
        &   & \ \ D_i\clor C_i\in I,\ i:1,\ldots,n\ \\
        &    & \ \ C=C'\clor C_1\clor\ldots\clor C_n\ \}.\\
\end{array}$$
\end{definition}
The operator $\dtp$ is monotonic and continuous on the lattice of 
interpretations ordered w.r.t. set inclusion.
Based on this property, the fixpoint semantics is defined as
$\dfs=\lfp(\dtp)=\dtp\!\!\uparrow_\omega$. 
As shown in \cite{MRL91},  
for all $C\in\dosp$ there exists $C'\in\dfs$ s.t. $C'$ implies $C$. 
Note that for two ground clauses $C$ and $C'$, 
$C$ implies $C'$ if and only if $C\subseteq C'$.
This suggests that interpretations can also be ordered w.r.t. subset inclusion 
for their elements, i.e., $I\sqsubseteq J$ if and only if for all $A\in I$ 
there exists $B\in J$ such that $B\subseteq A$ ($B$ implies $A$).
In the rest of the paper we will consider this latter ordering. 
\begin{example}
Consider the disjunctive program $P=\{r(a),\ p(X)\vee q(X)\leftarrow r(X)\}$
and the auxiliary predicate $t$. 
Then, $\dhbp=\{r(a),p(a),q(a),t(a),p(a)\vee r(a),p(a)\vee q(a),p(a)\vee q(a)\vee r(a),\ldots\}$.
Furthermore, the goal  
$G_0=\leftarrow (p(a)\vee q(a)\vee t(a))$ has the refutation
$G_0,G_1=\leftarrow  (p(a)\vee q(a)\vee t(a)\vee r(a)),G_2$
where
$G_2$ consists of the empty clause only.
The fixpoint semantics of $P$ is as follows $\dfs=\{r(a),\ p(a)\vee q(a)\}$.
Note that $p(a)\vee q(a)\vee t(a)$ is implied by  $p(a)\vee q(a)$.
\end{example}
We note that the definition of the $\dtp$ operator can be re-formulated in
such a way that its input and output domains contain multisets instead of sets
of atoms (i.e., we can consider interpretations which are sets of multisets of
atoms). In fact, we can always map a multiset to its {\em underlying set},
i.e., the set containing the elements with multiplicity greater than zero,
and, vice versa, a set can be viewed as a multiset in which each element has
multiplicity equal to one.
In the following we will assume that $\dtp$ is
defined on domains containing multisets. As the fixpoint operator for LO
is defined on the same kind of domains, this will
make the comparison between the two operators easier.
Furthermore, without loss of generality, we will make the assumption that in
clauses like 
$A_1\clor\ldots\clor A_n\cimp C_1\cand\ldots\cand C_m$,
the $A_i$'s are all {\em distinct}
and each $C_j$ consist of {\em distinct} atoms.
This will simplify the embedding of DLP clauses into linear logic. 
The previous definitions can be easily adapted.

Now, we give a closer look at the formal presentations of DLP and LO.
As said in the Introduction, we only need to consider a fragment of LO in
which connectives can not be arbitrarily nested in goals, like in DLP.
This fragment can be described by the following grammar:

$$
\begin{array}{l}
\formula{H}\ ::=\ \formula{A}_1\para\ \ldots\para\ \formula{A}_n\\
[\medskipamount]
\formula{D}\ ::=\ \formula{H}\ \leftlolli\ \formula{G}\ \ |\ \ \formula{D}\ \with\ \formula{D}\\
[\medskipamount]
\formula{G}\ ::=\ \formula{H}_1\with\ \ldots\with\ \formula{H}_n\ \ \ |\ \ \ \all
\end{array}
$$

where $\formula{A}_i$ is an atomic formula.
The comparison between the two languages is based on the idea that, to some
extent, linear connectives, i.e., additive conjunction $\with$ and
multiplicative disjunction $\para$, should behave like classical conjunction
$\cand$ and classical disjunction $\clor$. Actually, classical connectives
give rise to a fixpoint semantics for DLP which is a proper abstraction of
the semantics for LO. The translation between linear and classical connectives
is given via the following mapping $\trad{\cdot}$:
$$
\trad{F\clor G}=\trad{F}\para\trad{G},\
\trad{F\cand G}=\trad{F}\with\trad{G},\
\trad{F\cimp G}=\trad{F}\lollo\trad{G},\
\trad{\true}=\all.
$$
In order to make the comparison between DLP and LO more direct,
it is possible to present DLP by means of the following grammar:
$$
\begin{array}{l}
\formula{H}\ ::=\ \formula{A}_1\clor\ \ldots\clor\ \formula{A}_n\\
[\medskipamount]
\formula{D}\ ::=\ \formula{H}\ \cimp\ \formula{G}\ \ |\ \ \formula{D}\ \cand\ \formula{D}\\
[\medskipamount]
\formula{G}\ ::=\ \formula{H}_1\cand\ \ldots\cand\ \formula{H}_n\ \ \ |\ \ \ \true
\end{array}
$$
where $\formula{A}_i$ is an atomic formula.
A DLP program $P$ is now a $\formula{D}$-clause, whereas DLP goals
are represented (modulo `$\leftarrow$') as $\formula{G}$-formulas.
Here, we have introduced an explicit constant $\true$ for $true$ and we
have written {\em unit clauses} (i.e., with empty body) with the syntax
$ A_1\clor\ldots\clor A_n\cimp\true$. With these conventions, the grammars
for LO and DLP given above are exactly the same modulo the translation
$\trad{\cdot}$.
The definitions concerning the operational and fixpoint semantics for DLP
given previously can be adapted in a straightforward manner. 
The reader can also note that by definition of DLP program, the image of
$\trad{\cdot}$ returns a class of LO programs 
where both the head and the disjuncts in the body have no repeated occurrences
of the same atom.

We conclude this section by specializing our fixpoint semantics for LO,
given in Section \ref{nonground}, to the simpler fragment presented above.
We give the following definition for the $\lotp$ operator:
\begin{definition}[$\lotp$ operator]
Given an LO program $P$ and an interpretation $I$,
the operator $\lotp$ is defined as follows:
{\small $$\begin{array}{lll}
  \lotp(I) & = & \{\ms{H}+({\cal C}_1\bullet\ldots\bullet
  {\cal C}_n)\ |\ H\leftlolli D_1\with\ldots\with D_n\in P,
  \ \forall i=1,\ldots,n,\ \ms{D_i}+{\cal C}_i\in I\}\\
  & & \ \ \ \ \ \smallunion\ \ \ \ \ \{\ms{H}\ |\ H\lollo\all\in P\}
\end{array}$$}
\end{definition}
The operator $\lotp$ is monotonic and continuous over the lattice of Herbrand 
interpretations (ordered w.r.t. $\sqsubseteq$).
Thus, the fixpoint semantics of an LO-program $P$ is defined
as 
$$\lofs=\bigsqcup_{i=0}^\omega \lotp\!\!\uparrow_i.$$
A completeness result similar to that of Section \ref{nonground},
stating the equivalence between the operational and fixpoint semantics,
obviously holds for the fragment of LO considered here.

\section{DLP as Abstraction of LO}
\label{comparison}
The fixpoint semantics of LO allows us to investigate in more depth 
the relationships between LO and DLP.
For this purpose, we can employ the mathematical tools provided 
by {\em abstract interpretation} \cite{CC77}, and in particular the notion 
of {\em completeness} \cite{CC77,GR97b} that qualifies the precision of an abstraction.
Informally, the comparison between LO and DLP fixpoint semantics is based
on the abstraction that maps multisets into sets 
of atomic formulas (positive clauses). 
This abstraction induces a Galois connection between the semantic domains 
of DLP and LO. 
We prove that the fixpoint semantics of the
translation of 
an LO program in DLP is a correct abstraction of the fixpoint semantics of the
original LO program.
Furthermore, we show that this abstraction is not {\em fully complete}
with respect to LO semantics.
In a {\em fully complete} abstraction the result of interleaving 
the application of the abstract fixpoint operator with the abstraction
$\alpha$
coincides with the abstraction of the concrete fixpoint operator. 
For a {\em complete} abstraction, a similar relation holds for fixpoints, i.e., the fixpoint
of the abstract operator coincides with the abstraction of the fixpoint of the
concrete one.
We isolate an interesting class of LO programs 
for which we show that the property of completeness holds. In particular,
completeness holds if we forbid conjunctions in the body of
clauses. The resulting class of LO programs
is still very interesting, as it can be used to encode Petri Nets.

{\em Abstract Interpretation} \cite{CC77,CC79} is a classical framework for
semantics approximation which is used for the construction of semantics-based
program analysis algorithms. Given a semantics and an abstraction of the
language constructors and standard data, abstract interpretation determines an
abstract representation of the language which is, by construction, sound with
respect to the standard semantics. This new representation enables the
calculation of the abstract semantics in finite time, although it implies
some loss of precision. We recall here some key concepts in abstract
interpretation, which the reader can find in \cite{CC77, CC79, GR97b}.

Given a concrete semantics $\tuple{C,T_P}$, specified by a
{\em concrete domain} (complete lattice) $C$
and a (monotone) fixpoint operator $T_P:C\rightarrow C$, the abstract
semantics can be specified by an {\em abstract domain} $A$ and an abstract
fixpoint operator $\TPA$. In this context, program semantics is given by
$\lfp(T_P)$, and its abstraction is $\lfp(\TPA)$. The concrete and abstract
semantics $S=\tuple{C,T_P}$ and $S^\#=\tuple{A,\TPA}$ are related by a Galois
connection $\tuple{\alpha,C,A,\gamma}$, where $\alpha:C\rightarrow A$ and
$\gamma:A\rightarrow C$ are called {\em abstraction} and {\em concretization}
functions, respectively.
$S^\#$ is called a {\em sound} abstraction of $S$ if for all $P$,
$\alpha(\lfp(T_P))\leq_A \lfp(\TPA)$. This condition is implied by the
strongest property of {\em full soundness}, which requires that
$\alpha\compos T_P\leq_A\TPA\compos\alpha$. The notions of {\em completeness}
and {\em full completeness} are dual with respect to those of soundness.
Namely, $S^\#$ is a (fully) {\em complete} abstraction of $S$ if for all $P$,
($\TPA\compos\alpha\leq_A\alpha\compos T_P$)
$\lfp(\TPA)\leq_A\alpha(\lfp(T_P))$. Often, the notion of completeness is
assumed to include soundness (i.e., we impose '=' in the previous
equations).
It is well-known \cite{CC79} that the abstract domain $A$ induces a {\em best}
correct approximation of $T_P$, which is given by
$\alpha\compos T_P\compos\gamma$, and that it is possible to define a (fully)
complete abstract operator $\TPA$ if and only if the best correct
approximation is (fully) complete. It can be proved that for a fixed
concrete semantics, (full) completeness of an abstract interpretation only
depends on the choice of the abstract domain.
The problem of achieving a (fully) complete abstract interpretation starting
from a correct one, by either refining or simplifying the abstract domain,
is studied in \cite{GR97b}.
We conclude by observing that an equivalent presentation of abstract
interpretation is based on {\em closure operators} \cite{CC79}, i.e.
functions from a concrete domain $C$ to itself which are {\em monotone},
{\em idempotent} and {\em extensive}. This approach provides
independence from specific representations of abstract domain's objects
(the abstract domain is given by the image, i.e., the set of fixpoints, of
the closure operator).

We are now in the position of  connecting the LO (concrete) semantics 
with the DLP (abstract) semantics.
We define the abstract interpretation as a closure operator on the 
lattice $\calI$, the domain of LO interpretations of Definition \ref{interpretation_lattice}. 
In fact, as mentioned before, we can consider disjunctive interpretations as a subclass of $\calI$ 
(i.e., all sets in $\calI$).
We recall  that in $\calI$ the ordering of interpretations is defined as
follows: 
$I\sqsubseteq J$ iff for all $B\in I$ there exists $A\in J$ such that $A$ is a sub-multiset 
of $B$ (i.e., for disjunctive interpretations, $A\subseteq B$).
We give the following definitions.
\begin{definition} [Abstract Interpretation from LO to DLP]
The abstract interpretation is defined by the closure operator
$\alpha:\calI\rightarrow\calI$ such that
for every $I\in\calI$, $\alpha(I)=\{\alpha(\cal A)\ |\ \cal A\in I\}$, where
for a given multiset $\cal A$, $\alpha(\cal A)$ is the multiset such that
for every $i=1,\ldots,n$, $Occ_{\alpha(\cal A)}(a_i)=0$ if
$Occ_{\cal A}(a_i)=0$, $Occ_{\alpha(\cal A)}(a_i)=1$ otherwise (i.e., we
abstract a multiset with the corresponding set).
\end{definition}
\begin{definition}[Abstract semantics]
The abstract fixpoint semantics is given by $\lfp(\tpa)$, where the abstract
fixpoint operator $\tpa$ is defined as $(\alpha\compos\lotp)$.
\end{definition}
According to \cite{CC79}, $\alpha\compos\lotp$ is the best correct
approximation of the concrete fixpoint operator $\lotp$, for the fixed
abstraction $\alpha$. The abstraction $\alpha$, as said before, transforms
multisets into sets by forgetting multiple occurrences of atoms. 
It is not difficult to convince ourselves that $\tpa$ is indeed the $\dtp$
operator for disjunctive logic programs, provided that, as discussed
in Section \ref{dlplo}, we consider $\dtp$ defined over domains
containing multisets instead of sets (actually, we are identifying $\dtp$
input domain with the abstract domain which is given by the set of
fixpoints, i.e., the image, of the closure operator $\alpha$).
The operations $\bullet$ ({\em least upper bound} of multisets) and $+$
(multiset union) used in the definition of $\lotp$ are interchangeable
(because of the subsequent application of the operator $\alpha$)
and correspond to set (multiset) union in the definition of $\dtp$.
We have the following results.
\begin{proposition}[DLP is an abstraction of LO]
For every  DLP program $P$ and {\em disjunctive} (hence LO) interpretation $I$,
$\dtp(I)=\tpatr(I)$. 
\end{proposition}
\begin{proof}
By definitions.
\end{proof}
\begin{proposition}[DLP is a correct abstraction of LO]
For every LO program $P$,
the abstract semantics is a fully sound abstraction of the concrete semantics,
that is, for every interpretation $I$,
$\alpha(\lotp(I))\sqsubseteq\tpa(\alpha(I)).$
\end{proposition}
\begin{proof}
As $\tpa=\alpha\compos\lotp$ and $I\sqsubseteq\alpha(I)$, the proposition follows
by monotonicity of $\tpa$.
\end{proof}
The previous result implies {\em soundness}, i.e.
$\alpha(\lfp(\lotp))\sqsubseteq \lfp(\tpa)$.
The strong property of {\em full completeness} does not hold for the
abstraction. To see why, take as a counterexample the single clause
$a\leftlolli b$ and the interpretation 
$I$ with the single multiset $\{b,b\}$.
Then, $\alpha(\lotp(I))=\{\{a,b\}\}$, $\tpa(\alpha(I))=\{\{a\}\}$,
and $\tpa(\alpha(I))\not\sqsubseteq \alpha(\lotp(I))$.

We conclude this section showing that the abstraction is complete
for the subclass of LO programs whose clauses contain 
{\em at most} one conjunct in the body. We will address some applications
of this result in Section \ref{petri}.

(Note: the abstraction not being fully complete has a counterpart in 
the fact that in general $h>1$ in the following lemma, i.e., 
more than one step of $\lotp$ is necessary to simulate one step of $\tpa)$.)
\begin{lemma}
\label{complemma}
Let $P$ be an LO  program in which every clause has at most one conjunct in the
body (i.e., conjunction is forbidden), and $I,J$ two interpretations.
If $I\sqsubseteq\alpha(J)$ then there exists a natural number $h$ such
that $\alpha(\lotp(I))\sqsubseteq\alpha(\lotpo{h}(J))$.
\end{lemma}
\begin{proof}
Suppose $I\sqsubseteq\alpha(J)$ and $\calA\in\alpha(\lotp(I))$. Then
there exists $\calA'\in\lotp(I)$ s.t. $\calA=\alpha(\calA')$.
By definition of $\lotp$, there exists a clause $H\lollo D\in P$ (the case for
unit clauses is trivial)
s.t. $\ms{D}+\calC\in I$ and $\calA'=\ms{H}+\calC$. As
$I\sqsubseteq\alpha(J)$, we also have $\ms{D}+\calC\in\alpha(J)$,
which implies that there exists $\calK\in J$ s.t.
$\ms{D}+\calC=\alpha(\calK)$. Let
$p=min\{n\ |\ \calK\preccurlyeq(\ms{D}+\calC)^n\}$ (it is immediate to prove
that such a $p$ exists), and let $\calM=(\ms{D}+\calC)^p$.
We have that $\calK\preccurlyeq\calM$, therefore $\calM\in J$ (because
$\calK\in J$ and $J$ is upward-closed).
Now, $\calM=\ms{D}+(\calC+(\ms{D}+\calC)^{p-1})\in J$, and,
by definition of $\lotp$ ($H\lollo D\in P$),
$\ms{H}+\calC+(\ms{D}+\calC)^{p-1}\in\lotp(J)$. By repeatedly applying $\lotp$
(the proof is by induction on $p$) we get
$(\ms{H}+\calC)^p\in\lotpo{p}(J)$. Therefore
$\calA=\alpha(\ms{H}+\calC)=\alpha((\ms{H}+\calC)^p)\in\alpha(\lotpo{p}(J))$.
\end{proof}
\begin{proposition}
Let $P$ be an LO  program in which every clause has at most one conjunct in the
body. Then $\alpha(\lfp(\lotp))=\lfp(\tpa)$.
\end{proposition}
\begin{proof}
By a simple induction, using Lemma \ref{complemma}, we have that for every $k$
there exists $h$ s.t.
$\tpao{k}\sqsubseteq\alpha(\lotpo{h})$. Therefore
$\lfp(\tpa)\sqsubseteq\alpha(\lfp(\lotp))$.
\end{proof}
The class of LO programs with one conjunct in the body is still very
interesting. Below, we show how this result could be exploited to study
reachability problems in Petri Nets.
\section{Other Applications: Relation with Petri Nets}
\label{petri}
As shown in the proof of Proposition \ref{undecidable}, the class of propositional 
LO programs with one conjunct in the body is equivalent to VAS, i.e., to Petri Nets. 
Intuitively, a multiset rewriting rule can be used to describe the effect of firing 
a Petri Net transition.
For instance, the clause $a~\para~b~\para~b~\lollo~c~\para~c$
can be interpreted as the Petri Net transition that removes one token from 
place $a$, two tokens from place $b$, and adds two tokens to place $c$.  
As a consequence, a (possibly infinite) execution (sequence of goals $G_0,G_1,\ldots$) 
of a restricted LO program denotes an execution of the corresponding Petri Net. 
The initial goal $G_0$ can be viewed as the initial marking of the Petri Net.
Consider now the fact $F~\equiv~c\leftlolli\all$, and let $G_0$ be the goal $a\para a\para b$.
Then, the sequent $P\cup F\Rightarrow G_0$ is provable in LO
if and only if there exists a reachable marking having {\em at least} 
one token in place $c$.
In other words, the fact $F$ can be used to implicitly represent an infinite set of 
markings (its upward closure) of the corresponding Petri Net. 
Our bottom-up semantics can be use to effectively compute the set $Pre^*(F)$ 
(using the terminology of \cite{ACJT96}) of markings that can reach a marking in 
the denotation of $F$.

This idea can be used to verify {\em safety properties} of concurrent systems.
A safety property $S$ can be viewed as a set of {\em good} states (markings) of a given concurrent 
system (Petri Net).
The system satisfies the property if the set of states that are reachable from the initial 
state $G_0$ are all contained in $S$.
Symmetrically, the set $\neg S$ represents the set of {\em bad} states. 
Thus, the systems can  be proved correct by showing that $Pre^*(\neg S)$ does not contain 
the initial state $G_0$, i.e., by applying the bottom-up algorithm starting from 
a fact denoting $\neg S$.
It is interesting to note that in many real examples $\neg S$ is indeed an upward closed set 
of states (e.g.  the set of states where there are 
{\em at least two processes in the critical section} are the the typical bad states of 
a mutual exclusion algorithm).
In general, the complexity of computing $Pre^*(F)$, for some $F$, can be very high.
However, the results of Section \ref{comparison} show that the fixpoint semantics of DLP can 
be used to approximate the set $Pre^*(F)$.
Completeness implies that all properties that are preserved by the abstraction 
can be checked equivalently over the concrete and the abstract domain.
In our setting the kind of properties that satisfy this requirement can be informally 
characterized as `at least one'-properties (e.g. {\em  is there at least one token in 
place P in a reachable marking?}).
This kind of properties can be used to check `mutual exclusion' for a concurrent system 
represented via a Petri Net.
Suppose we want to prove that a system ensures mutual exclusion 
for two processes represented via a Petri Net. Process $p_i$ is in the critical section 
whenever a token is in a special place $cs_i$ for $i:1,2$. 
Violations of mutual exclusion are expressed as the set of states with {\em at least} one token in place 
$cs_1$ and one token in state $cs_2$.
Thus, the fixpoint semantics of the DLP program obtained as translation 
of the Petri Net (LO program) union the fact $cs_1\vee cs_2$ is an abstraction 
of the set of backward reachable states. 
We obtain a full-test for mutual exclusion properties, whenever the initial states 
can be expressed again as {\em at least one} properties (i.e., whenever membership of the 
initial states in the set of abstract reachable states can be determined exactly).
\section{Related Works}
\label{related}
Our work is inspired to the general decidability results for infinite-state systems
based on the theory of well-quasi orderings given in \cite{ACJT96,FS01}. 
In fact, the construction of the least fixpoint of $\STP$ and $\STPo$ can be viewed 
as an instance of the {\em backward reachability} algorithm for transition systems 
presented in \cite{ACJT96}.
Differently from \cite{ACJT96,FS01},  we need to add special rules 
(via the satisfiability relation $\asat$) to handle formulas with the connectives 
$\with$, $\all$ and $\one$. 

Other sources of inspiration come from linear logic programming.
In \cite{HW98}, the authors present an abstract deductive system
for the bottom-up evaluation of linear logic programs. 
The {\em left introduction} rules plus {\em weakening} and {\em cut} are used to compute 
the logical consequences of a given formula.
The satisfiability relations we use in the definition of the fixpoint 
operators correspond to top-down steps within their bottom-up evaluation 
scheme.
The framework is given for a more general fragment than {LO}. 
However, they do not provide an {\em effective} 
fixpoint operator as we did in the case of {LO} and \LOonews, and
they do not discuss computability issues for the resulting 
derivability relation.  

In \cite{APC97}, Andreoli, Pareschi and Castagnetti present a partial 
evaluation scheme for propositional {LO} (i.e without $\one$).
Given an initial goal $G$, they use a construction similar to Karp and Miller's
{\em coverability tree} \cite{KM69} for Petri Nets to build a finite representation 
of a proof tree for $G$. 
During the {\em top-down} construction of the graph for $G$,
they apply in fact a {\em generalization step} that works as follows.
If a goal, say $\calB$, that has to be proved is subsumed by a node already visited, 
say $\calA$, (i.e., $\calB=\calA+\calA'$), then the part of proof tree between the two 
goals is replaced by a proof tree for $\calA+(\calA')^*$;
$\calA+(\calA')^*$ is a finite representation of the union of $\calA$ with 
the closure of $\calA'$.
They use Dickson's Lemma to show that the construction always terminates.
In the case of {LO}, the main difference with our approach is that we give 
a goal independent {\em bottom-up} algorithm.
Technically, another difference is that in our fixpoint semantics we do not 
need any {\em generalization} step. In fact, in our setting the 
computation starts directly from (a representation of) {\em upward-closed} sets of contexts. 
This simplifies the computation as shown in Example \ref{example} 
(we only need to test $\preccurlyeq$).
Finally,  differently from \cite{APC97}, in this paper we have given also a formal 
semantics for the extension of {LO} with the constant $\one$.

The partial evaluation scheme of \cite{APC97} is aimed at com\-pile-time
optimizations of abstractions of {\em LinLog} programs.
Another example of analysis of concurrent languages based on linear logic 
is given in \cite{KNY95}, where the authors present a type inference
procedure that returns an approximation of the number of messages 
exchanged by HACL processes.

In \cite{Cer95} Cervesato shows how to encode Petri Nets in {LO}, 
Lolli and Forum by exploiting the different features of these languages.  
We used some of these ideas to prove Proposition \ref{undecidable}.

Finally, we have discussed the similarities between our semantics and
the bottom-up semantics for 
Disjunctive Logic Programming of Minker, Rajasekar and Lobo \cite{MRL91}.
In a disjunctive logic program, the head of a clause is a disjunction
of atomic
formulas, whereas the body is a conjunction of atomic formulas.
In the semantics of \cite{MRL91} interpretations are collections of
{\em sets} (disjunctions)
of atomic formulas. Only minimal (w.r.t. set inclusion) sets are kept at each
fixpoint iteration.
In contrast, in our setting we need to consider collections of {\em multisets}
of formulas.
Therefore, in the propositional case in order to prove the 
convergence of the fixpoint iteration, we need an argument (Dickson's lemma) 
stronger than the finiteness of the {\em extended} Herbrand base of
\cite{MRL91} (collection of all minimal sets).
\section{Conclusions and Future Work}
\label{conclusions}
In this paper we have defined a bottom-up semantics for the fragment 
of LinLog \cite{And92} consisting of  the language {LO} \cite{AP91a} 
enriched with the constant $\one$.
In the propositional case, we have shown that without $\one$ 
the fixpoint semantics is finitely computable. 
Our fixpoint operator is defined over constraints and gives us an effective 
way to evaluate bottom-up (abstractions of) linear logic programs. 
To conclude, let us discuss the directions of research that we find more promising.
\smallskip\\
{\em Linear Logic Programming.}
It would be interesting to extend the techniques we presented in this paper 
to larger fragments of linear logic. 
In particular, it would be interesting to define a  bottom-up evaluation 
for languages like Lolli \cite{HM94} and
Lygon \cite{HP94}, and to study techniques for first-order 
formulation for all these languages.
An extension of the present framework to the first-order case should also
take into account the so-called {\em S-semantics}
 \cite{FLMP93,BGLM94}, in order to model
observables like {\em computed answer substitutions} and to cope with issues
like {\em compositional} semantics. Concerning LO, we would also like to
look at the connection with the so-called Chemical Abstract Machine
metaphor \cite{ALPT93}.
\smallskip\\
{\em Verification.}
In \cite{DP99}, the authors show that properties of
concurrent systems expressed in {\em temporal logic} can be defined in terms 
of fixpoint semantics of a logic program  that encodes the transition system 
of a concurrent  system.
In \cite{DP99}, synchronization between processes is achieved via 
{\em shared variables}, whereas in linear logic synchronization can be expressed 
via multiple headed clauses. 
Thus, our semantics might be a first step towards the extension of 
the metaphor of \cite{DP99} to concurrent systems in which  synchronization 
is expressed at the logical level (see Section \ref{petri}).
The other way round, through the connection between semantics and verification,
techniques  used for infinite-state systems with integer variables 
(see e.g. \cite{DP99,BGP97,HHW97}) can be re-used in order to  compute a 
static analysis of linear logic programs.
\smallskip\\
{\em Proof Theory.}
The connection we establish in this paper indicates a potential connection
between the general decidability results for infinite-state systems of \cite{ACJT96,FS01} 
and provability in {\em sub-structural} logics like {LO} and {\em affine} linear logic
\cite{Kop95}.
Viewing the provability relation as a transition relation, it might be possible to find
a notion of {\em well-structured} proof system (paraphrasing the notion of 
{\em well-structured} transition systems of \cite{ACJT96,FS01}), i.e., 
a general notion of provability that ensures the termination of the bottom-up
generation of valid formulas.
\smallskip\\
{\em Relations between DLP and LO.}
We hope that our study will give rise to new ideas for the analysis of
LO programs.
As an example, it could be interesting to study weak notions
of negation for LO that are based on the  negation of DLP.
Moreover, we can use DLP operational and fixpoint semantics to analyze
Petri Nets,
given that the abstraction is complete in this case.
Finally, there are still some open questions concerning the relation 
between DLP and LO in the setting of abstract interpretation.
In particular, we would like to study the notion of completeness for the
general class of LO programs (we remark that the example in \cite{BDM00b}
showing incompleteness was wrong). We would also like to analyze in more detail
the connection between the notion of (full) completeness of the
abstraction and proof-theoretic properties of provability in
sub-structural logics, which has been only partly addressed in
\cite{BDM00b}.

\section*{Acknowledgments}
The authors would like to thank Maurizio Gabbrielli for his support, 
Jean-Marc Andreoli for useful comments on the use of LO, 
and the anonymous reviewers for  providing us suggestions and references
that helped us to improve the presentation of the paper.

\bibliographystyle{tlp}
\bibliography{biblio}

\end{document}